\documentclass[preprint,amssymb,showpacs,supersriptaddress,floatfix]{revtex4-1}
\usepackage{graphicx}
\usepackage{dcolumn}
\usepackage{bm}
\usepackage{color}

\usepackage[utf8]{inputenc}
\usepackage[T1]{fontenc}
\usepackage{mathptmx}
\usepackage{etoolbox}

\begin{document}

\preprint{AIP/123-QED}

\title{Oxygen vacancy dynamics in Pt/TiO$_x$/TaO$_y$/Pt memristors: exchange with the environment and internal electromigration}



\author{Rodrigo Leal Martir$^{1,2}$, María José Sánchez$^{2,3}$, Myriam Aguirre$^{4,5}$, Walter Quiñonez$^{1,2}$ Cristian Ferreyra$^{1,2}$, Carlos Acha$^{6}$, Jerome Lecourt$^7$, Ulrike Lüders$^7$, Diego Rubi   $^{1,2}$}


\affiliation{$^{1}$Departamento de Micro y Nanotecnologías, Centro Atómico Constituyentes, Comisión Nacional de Energía Atómica, Gral Paz. 1499 (1650), San Martín, Argentina\\
$^{2}$ Instituto de Nanociencia y Nanotecnología (INN), CONICET-CNEA, Buenos Aires and Bariloche, Argentina \\ $^{3}$ Centro Atómico Bariloche and Instituto Balseiro (Universidad Nacional de Cuyo), 8400 San Carlos de Bariloche, Río Negro, Argentina \\ $^{4}$ Instituto de Nanociencia y Materiales de Aragón (INMA-CSIC) and  Dpto. de Física de la Materia Condensada, Universidad de Zaragoza, Spain \\ $^{5}$ Laboratorio de Microscopías Avanzadas, Edificio I+D, Campus Rio Ebro C/Mariano Esquillor s/n, 50018 Zaragoza, Spain. \\ $^{6}$ Depto. de Física, FCEyN, Universidad de Buenos Aires and IFIBA, UBA-CONICET, Pab I, Ciudad
Universitaria, Buenos Aires (1428), Argentina  \\  $^{7}$ CRISMAT, CNRS UMR 6508, ENSICAEN, 6 Boulevard Maréchal Juin, F-14050 Caen Cedex 4, France }

\date{\today}

\begin{abstract}


Memristors are expected to be one of the key building blocks for the development of new bio-inspired nanoelectronics. Memristive effects in transition metal oxides are usually linked to the electromigration at the nanoscale of charged oxygen vacancies (OV). In this paper we address, for Pt/TiO$_x$/TaO$_y$/Pt devices, the exchange of OV  between the device and the environment upon the application of electrical stress. From a combination of experiments and theoretical simulations we determine that both TiO$_x$ and TaO$_y$ layers  oxidize, via environmental oxygen uptake, during the electroforming process. 
Once the memristive effect is stabilized (post-forming behavior) our results suggest that oxygen exchange with the environment is suppressed and the OV dynamics that drives the memristive behavior is restricted to an internal electromigration between TiO$_x$ and TaO$_y$ layers. Our work provides relevant information for the design of reliable binary oxide memristive devices.

\end{abstract}

\maketitle

\section{Introduction}
Memristive systems -defined as metal/insulator/metal structures able to switch between different resistive states upon the application of external electrical stimuli \cite{saw_2008,iel_2016}- are expected to be one of the key building blocks for the development of new neuromorphic hardware \cite{yu_2017}, intended to outperform current software-based machine learning algorithms running on computers with the Von Neumann architecture \cite{trave_2015} . Memristive mechanisms strongly rely on the presence and electromigration of defects \cite{saw_2008,yang_2008}; in the case of oxides, these defects are usually the ubiquitous charged OV \cite{gun20}. Typically, the electromigration of OV can lead into the formation and disruption of conducting nanofilaments or to the modulation of the resistance of Schottky metal/oxide interfaces \cite{saw_2008}. It has been reported that both mechanisms could coexist for single devices, being possible to select one or the other by controlling external stimuli parameters such as the compliance current programmed during the transition from high to low resistance states \cite{rubi_2013} or other device operation conditions \cite{muenster_2010}.


Among single oxides, TaO$_y$ presents a high potential to be implemented in memristive systems with neuromorphic behavior. This is based on its CMOS compatibility -which would ease the integration with standard electronics-, analog response - in order to mimic the adaptable synaptic weights of biological synapses- \cite{ryu_2011,wang_2015, choi_2018}, very high endurance (from 10$^{10}$ to 10$^{12}$ cycles) \cite{yang_2010, lee_2018}, ultrafast switching time ($\approx$ 10 ps) \cite{bottger_2020}, large ON-OFF ratio ($\approx$ 10$^6$) \cite{shi_2017} and ultra-low power operation ($\approx$ 60 fJ/bit) \cite{shi_2017}. The already reported memristive mechanisms in TaO$_y$-based devices include the formation of conducting nanofilaments \cite{ryu_2011, wedig_2015, park_2015, heisig_2022}, the modulation of energy barriers present at  metal-oxide interfaces \cite{hsu_2014, wang_2015} or the OV exchange between TaO$_y$/TaO$_h$ bilayers (with different degree of oxidation) \cite{yang_2012, lee_2018, ferreyra_2020}. In the latter case, it is usually assumed that the more reduced layer acts as OV source/sink that eases the reduction/oxidation of the more oxidized one that drives the resistance changes -usually Ta$_2$O$_5$- \cite{lee_2018}.

Many mechanisms proposed to describe the memristive behavior of oxide-based devices have usually assumed that OV dynamics takes place internally between different layers or zones of the device \cite{nian_2007, rozenberg_2010}; in other words, the total amount of OV present in the device is considered as a constant. More recently, it has been experimentally shown for different oxide-based memristors that molecular oxygen transfer across oxide-metal interfaces \cite{zhang_2018,siegel_2021}, which could eventually lead to oxygen exchange with the environment, could be a non-negligible effect and must be taken into account to properly describe the memristive effect. Advanced characterization tools such as in-operando transmission electron microscopy \cite{cooper_2017} or secondary mass ion spectrometry \cite{cox_2021} were used to get evidence on this. It has been proposed that moisture seems to play a key role in providing oxygen to the device oxidation \cite{messer_2015,heisig_2018,5306148}. We also notice that the incorporation of protons to the device was also proposed to affect the device electrical behavior in the case of cationic resistive switches \cite{Tsuruoka_2016}. The influence of ambient conditions on the memristive response of the device is therefore not a trivial issue for the technological applications of these systems, and it needs to be fully understood and controlled in order to develop strategies -such as, for example, a proper encapsulation of the device if necessary- to warrant a reliable memristive behavior.

In this paper, we address, from a combination of experiments and theoretical simulations, the memristive response of Pt/TiO$_x$/TaO$_y$/Pt heterostructures, making focus on the presence of oxygen exchange with the environment. Our findings indicate that the electroforming process is accompanied by a strong oxygen uptake from the ambient -which oxidizes both TiO$_x$ and TaO$_y$ layers-, but after the memristive cycling is stabilized oxygen exchange with the environment is spontaneously suppressed and OV dynamics is restricted to an internal exchange between TiO$_x$ and TaO$_y$ layers.

\section{Methods}

We have grown by pulsed laser deposition TiO$_x$/TaO$_y$ bilayers on top of platinized silicon substrates. The depositions were made at room temperature and at oxygen pressures of 0.01 and 0.1 mbar, respectively. Top Pt electrodes were microfabricated by a combination of sputtering and optical lithography. Electrical characterization was performed with a source measure unit Keithley 2612B hooked to a commercial probe station. High resolution Scanning Transmission Electron Microscopy with a High Angular Annular Dark Field
Detector (STEM-HAADF) was performed using a FEI Titan G2
microscope with a probe corrector (60–300 keV). In situ chemical
analysis was performed by Energy Dispersive Spectroscopy (EDS). Samples for TEM
were prepared by Focused Ion Beam (FIB) in a Helios 650 dual beam equipment.

\section{Device electroforming}

Fig. 1(a) shows a STEM-HAADF cross-section corresponding to a virgin Pt/TiO$_x$/TaO$_y$/Pt heterostructure, before the application of any voltage stress. The STEM-HAADF image suggests that TiO$_x$ and TaO$_y$ thicknesses are 81.5 nm and 10.5 nm, respectively. EDS linescans, shown in Fig. 1(b), indicate composition gradients for both oxide layers: the Ti oxide chemistry goes from TiO$_{1.22}$ in the region close to the top Pt electrode to TiO$_{2.24}$ close to the interface with Ta oxide. On the other hand, the Ta oxide layer displays a stochiometry ranging from TaO$_{2.33}$ close the interface with the Ti oxide to TaO$_{3}$ close to the bottom Pt electrode. We notice that memristors with graded chemical composition and reliable behavior were reported in the literature \cite{wang_2018,gul_2022}. EDS line scans also show that the TiO$_x$/TaO$_y$ interface is not sharp but it displays a zone ($\approx$ 10 nm) of Ti and Ta intermixing. Fast Fourier Transforms (FFT) performed in both oxide layers (not shown here) show the absence of diffraction poles, evidencing their amorphous character. In addition, the top Pt electrode shows the presence of columns and grain boundaries, which could behave eventually as fast paths for oxygen migration in and out of the device \cite{zurhelle_2022}.

\begin{figure} 
\centering
\includegraphics[width=1\linewidth]{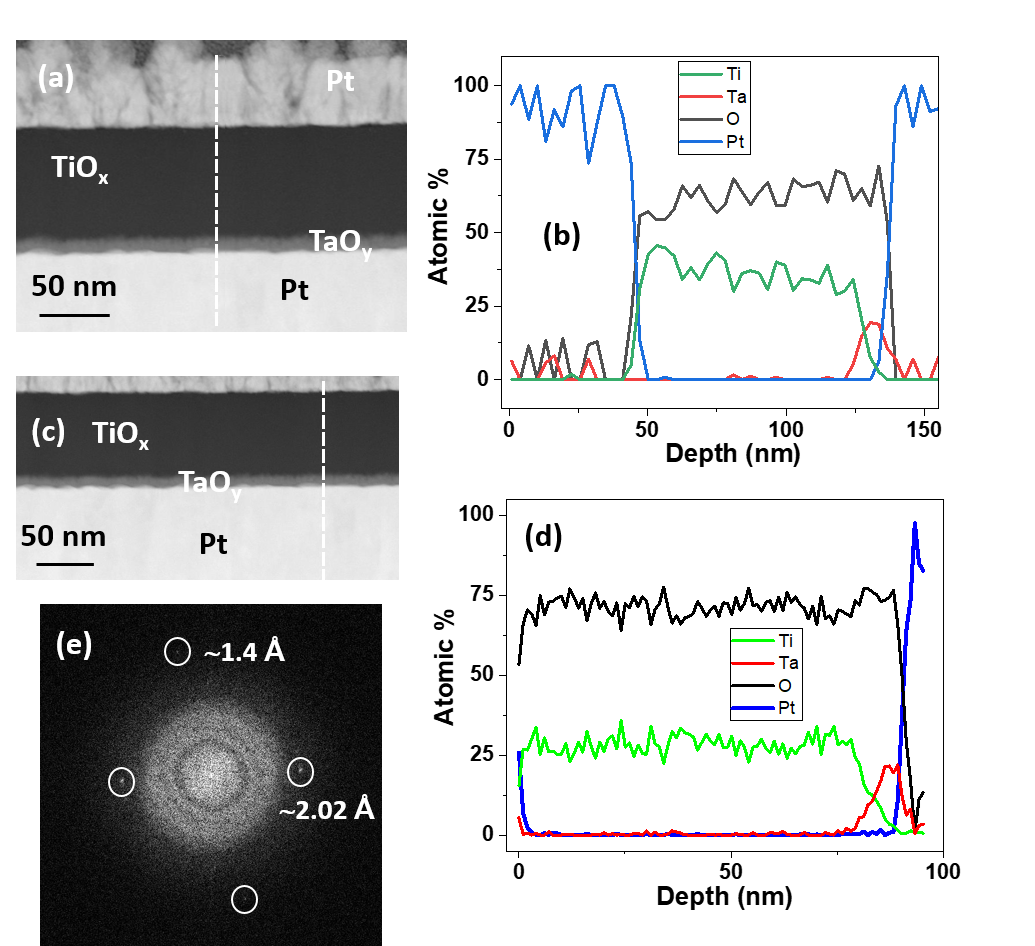}
\caption{(a) STEM-HAADF cross-section corresponding to a pristine Pt/TiO$_x$/TaO$_y$/Pt device; (b) EDS line-scans corresponding to the cross-section shown in (a), as shown with a dashed white line. Ti, Ta, O and Pt species are quantified. The scans start at the Pt top electrode and end at the bottom Pt electrode; (c) STEM-HAADF cross-section corresponding to a electroformed Pt/TiO$_x$/TaO$_y$/Pt device; (d) EDS line-scans corresponding to the formed device, shown in (c). The scans, shown with a dashed white line in (c), start at the Pt top electrode and end at the bottom Pt electrode; (e) FFT corresponding to the TiO$_x$ layer of the formed device. The appearance of faint diffraction poles -estimated interplanar distances are shown in the figure- indicates the formation of nanocrystallites but does not allow a precise identification of the crystalline phase.}
\label{Figure1}
\end{figure}

The forming process, shown in Fig. 2 for a $\approx 3x10^3$ $\mu$m$^2$ device, started with the application of an initial pulsed ramp consisting in voltage pulses (10ms wide) of increasing amplitudes from 0 V to V$_{FO}$ $\approx$ -5.5 V (see Fig. 2(b)). After each pulse, we measured the remanent resistance by applying a small voltage  of 100 mV. This initial stimulus produced a spike-like change in the device resistance, as it is observed in Fig. 2(c), from a virgin state of $\approx$100$ \Omega$ to $\approx$80 k$\Omega$ and then to $\approx$8 k$\Omega$. Afterwards, symmetric pulsed ramps with V$_{MIN}$ = -2 and -2.5 V and V$_{MAX}$ = 2 and 2.5 V, respectively, were applied -notice that -V$_{MIN}$, V$_{MAX}$ < -V$_{FO}$-, which produced a progressive resistance recovery until a stable resistive switching effect  between $\approx$10 k$\Omega$ and $\approx$14 k$\Omega$ was found, as it is displayed in Fig. 2(d) .


Figs. 1(c) and 1(d) show a STEM-HAADF cross-section of a formed device and the corresponding EDS linescans, respectively. It is found that, upon forming, both Ti and Ta oxide layers become more oxidized in relation to the virgin device: the Ti oxide layer displays an uniform TiO$_{2.3}$ stochiometry while the Ta oxide layer displays a TaO$_{3.4}$ stochiometry, also uniform in thickness. We notice that both layers display higher oxygen content than the standard (and stable) TiO$_{2}$ and Ta$_{2}$O$_{5}$ phases, as has been reported for TaO$_{y}$ \cite{tsuchiya_2011} and related to the absorption of environmental water molecules and the formation of Ta-O-O-H bonds by means of a protonation reaction \cite{manne_2016}. Also, FFT performed from the STEM-HAADF cross section show, for the formed device, the appearance of faint diffraction poles, indicating the formation of nanocrystallites of Ti and Ta oxides (see Fig. 1(e) for the case of Ti oxide). This crystallization process is likely related to the presence of thermal effects -via Joule heating- \cite{carta_2016,lederer_2021} during the initial stage of the electroforming process, and we associate it to the resistance spike we described before. 

The  experiments described above show that environmental oxygen -and eventually protons- are incorporated to the device upon forming. However, it is unclear if this interaction with the ambient is maintained once the resistive switching effect becomes stable. In order to tackle this issue, we have performed numerical simulations using the Voltage Enhance OV drift model (VEOV) \cite{rozenberg_2010, ghenzi_2013}, adapted here to describe the oxidation process during forming and the subsequent remanent resistance vs. voltage cycles in the Pt/TiO$_x$/TaO$_y$/Pt system.

\section{\label{sec:2}Simulating the electroforming process}

The VEOV model  simulates the migration of OV, ubiquitous in transition metal oxides,  due to an applied external electrical stimulus and  has been extensively 
 employed to unveil the memristive response of several  oxide based devices, even ferroelectrics  and topotactic manganites \cite{ghenzi_2014,Acevedo_2017, ferreyra_2020,ferreyra_2020_2,acevedo_2020,acevedo_2022}.
The  key ingredients of the model are i) the dependence of the resistivity of an oxide  on  its local oxygen stoichiometry and ii) the strong electric fields that develop close to the electrode(s)/oxide  interface(s) -usually forming Schottky barriers- and/or  at the interface   between different oxides composing   the device. Under an external voltage, OV   electromigrate back and forth depending on the polarity of the applied stimulus along  nanoscale regions close to where strong electric fields develop, with  the concomitant change in    the device resistance. In particular, TiO$_x$ and TaO$_y$ behave as an n-type semiconductors  in which  OV  are electron donors.

 To have further  insight into  the model details, Fig. 2(a) shows a sketch of the  present device where we have defined the memristive active regions relevant for the simulations.
The left (L) region  comprises the interface  Pt/TiO$_{x}$ while  the right (R) one represents the  interface   TaO$_y$/Pt. The central (C)  region   
mainly comprises the interface TiO$_{x}$/TaO$_y$, where our TEM experiments evidenced some Ti and Ta intermixing (recall Fig. 1).
Due to the largest thickness of the TiO$_{x}$ layer in our devices, we assume that the L region is larger than the  C and R  respectively,  the latter two including all the   TaO$_{y}$ layer. On the other hand, 
the  remaining  TiO$_{x}$ at the right of the L zone represents an inert bus zone  for the transfer of 
OV that, as we will show below,  essentially does not participate in the RS effect.
 
For the simulations we define  a  1D chain of  $N= NL+NC+NR+ NB$ total sites, where 
the first  $NL$ sites correspond to the  L layer,  $NC$ sites to  the central C layer 
and  $NR$ sites  to the  R layer, respectively. In addition $NB$ sites are assigned to the 
bus region in the TiO$_x$. Taking into account  the previous device description (recall the  STEM-HAADF cross-section displayed in Fig. 1),  we consider $NB> NL > NC>NR$ (see Table 1 for values of the parameters employed in the simulations).
Each site $i$ represents a domain of (sub)nanoscopic dimensions  characterized 
by its resisitivity $\rho_{i}=\rho_{0} (1 - A_{i} \delta_{i})$ \cite{ghenzi_2013} that decreases  with $\delta_{i}$, the local density of OV.
We define  ${\rho_{0}}$ as  the  residual resistivity  for negligible OV concentration and,  following the reported  resistivities for  TiO$_{x}$ and TaO$_{y}$ \cite{tseng_1999, arif_2017} we take different values of ${\rho_{0}}$ in each oxide, accordingly (see also Table 1). 
The coefficients  $A_{i}$ characterize the different interfaces and 
can be taken either   smoothly  dependent on the site position 
or as constants (as we do for simplicity), without affecting the qualitative behaviour of the simulated  results.
In all the cases we have $ A_{i} \delta_{i} < 1 \; \forall i$.

\begin{table}

\begin{center}\begin{tabular}[c]{|c | c | c | c | c|}
\hline
Region & Sites &V$_\alpha$ &$\rho_0$ (k$\Omega$)&A$_i$\\		        
\hline
L  & 60 &0.005 &48.1 &1.18\\	
\hline
C & 39   &0.007  &111.1 &0.56\\			    
\hline
R & 12   & 0.007  &111.1&0.28\\			    
\hline

\end{tabular}\end{center}
	\caption{Parameters employed in the numerical simulations of OV dynamics.
	The activation voltages $V_\alpha$ are in units of the thermal energy $K_B T$.
	The reservoir´s activation voltage was taken $V_{RE}=0.16$. The bus zone B has
	$N_B=297$ sites and same parameters than the L zone. 
	} 
\end{table}

The total resistivity of the system can be computed as
$\rho = \sum_{i=1}^{N}{\rho_{i}}$ and,  as we are considering a 1D model, the resistance
$R$ can be trivially computed from $\rho$ through a length scaling factor.

To account for the absorption of oxygen  observed experimentally during the forming process (see Fig. 1(d)), we assume that the sample can exchange OV with an external reservoir and thus the total number of OV in the sample is not conserved during the application of electrical stress. This is a  new key ingredient  that settles a difference with previous studies \cite{ghenzi_2013,ferreyra_2020,ferreyra_2020_2}, in which any possible exchange of OV between the sample and the ambient was neglected  in the VEOV model.

Following this line of reasoning, a  net decrease in the sample OV content will be interpreted as an oxidation process (a net uptake of oxygen). Although this might be a oversimplified assumption, as the net uptake of oxygen could be concomitant with other effects such as proton incorporation, as we mentioned before, it allows capturing non-trivial characteristics of the experimental forming process, as we will describe below.

We simulate the external reservoir  as a region in contact 
with the L zone that   can allocate  an arbitrarily large number of OV.
Notice that as reported in Figs. 1(c) and 1(d), the  uptake of oxygen is more favoured at the oxide layer   close to the Pt  top electrode and thus in the simulations we consider  that the interchange of OV is  through  the L zone and the reservoir. 

We emphasize that  the electroforming is a complex  out-of-equilibrium process under which structural changes, like crystallization, can additionally contribute  to the change of  the device resistance. We assume that: i) the  crystallization process is concomitant with the resistance spike we observed at the first stage of the electroforming process, and it is finished afterwards ii) after crystallization, the device still presents a large number of OV which are progressively filled during the next electroforming steps. Our simulations start at ii) and describe the resistance evolution assuming that no further structural changes occur.


In Fig. 2(b) we show the experimental voltage protocol $V(t)$ during the forming process together with the  $V_s (t)$ employed for the simulation, which follows quite well the experimental one.
Given a value of $V_s (t)$, the OV density at each site \textit{i}  is updated for each simulation step according to the  rate probability  $p_{ij} = \delta_i (1-\delta_j) \exp(-V_{\alpha} + \Delta V_{i})$ \cite{rozenberg_2010} for a transfer 
from site \textit{i} to a nearest neighbor \textit{j}= \textit{i} $ \pm 1$. 
Notice that $p_{ij}$  is proportional to the  OV density at site \textit{i} and to the available OV density  at the neighbour site \textit{j}.

In the  Arrhenius factor, $\exp(-V_{\alpha} + \Delta V_{i})$, $\Delta V_i$ is the 
local potential drop  at site \textit{i} defined as ${\Delta V}_i (t) = V_{i}(t) - V_{i-1}(t)$ with $V_i(t) =  V_s (t) \rho_{i} / \rho$ and $V_\alpha$ the activation energy for vacancy diffusion in the absence of external stimulus. The  values of  $V_\alpha = V_{L}, V_{B}, V_{C}$ and $V_R$ for the L, B, C and R layers are given in Table 1. In all the calculations, the energy scales are taken in units of the thermal energy $k_{B}T$.

As we mentioned, the reservoir is modelled by a region external to the sample which can allocate a large amount of OV. Thus following the usual statistical assumption, once OV are injected into the  reservoir they tend to remain in it.
To accomplish this,  we consider for the reservoir a transfer rate $p_{ij}$ with an  activation energy  $V_\alpha= V_{RE}\gg V_L$. In addition  we  do not include the external voltage  in the reservoir.

At each simulation time step $t_k$, we compute the local voltage profile $V_i(t_k)$, the local voltage drops ${\Delta V}_i (t_k)$ and   employing the probability rates $p_{ij}$ we obtain the transfers between nearest neighboring sites. 
Afterwards, the values $\delta_i(t_k)$ are updated to a new set of densities $\delta_i(t_{k+1})$, with which we compute at time $t_{k+1}$, the local resistivities  $\rho_i(t_{k+1})$, 
the local voltage drops under the applied voltage $V_s (t_{k+1})$, and  finally  the total resistivity $\rho(t_{k+1})$,  to start the next simulation step at $t_{k+1}$. 

\begin{figure*} 
\centering
\includegraphics[width=1\linewidth]{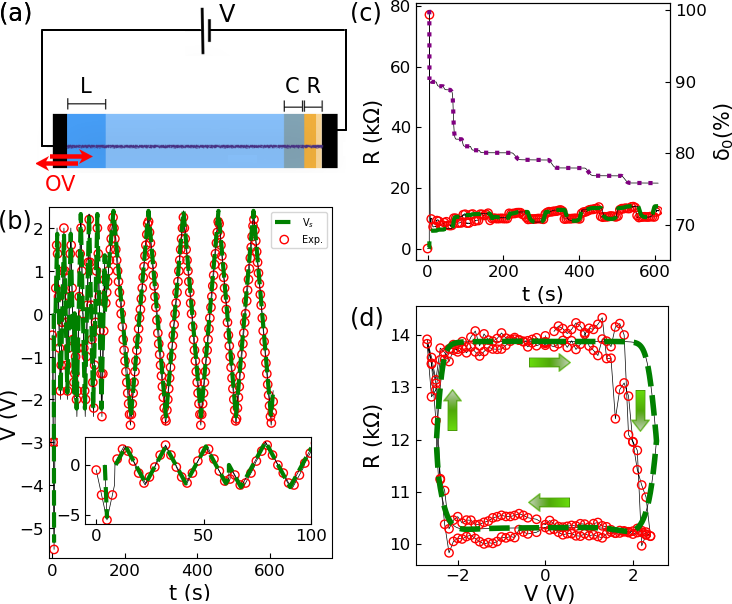}
\caption{(a) Scheme  of the device assumed for the simulations. The different active zones -L,R and C-  are indicated. The red arrows sketch  the exchange of OV with the ambient during the electroforming process; (b) Simulated (green dashed) and experimental (red circles) electroforming  voltage protocols. The inset shows a blow-up of the main panel for times between 0 and 100 s; (c) Left axis: resistance evolution with time during the  electroforming process of a $\approx 3x10^3$ $\mu$m$^2$ Pt/TiO$_x$/TaO$_y$/Pt device. The red (green) curve  corresponds to the experimental (simulated) remanent resistance values. Right axis: simulated OV percentage remaining  in the TiO$_x$/TaO$_y$ bilayer during the electroforming  process, as a function of time; (d) Stable remanent resistance vs. voltage cycle measured experimentally (red) and simulated (green), for the same device.}
\label{Figura2}
\end{figure*}

We consider an initial OV configuration   consistent with  
the value of $\approx$8k$\Omega$  attained in the experiment after the resistance spike observed in the first stage of the electroforming process.
Therefore  we start the simulations at time $0^{+}$, immediately after the application of a post-resistance spike short positive pulse (that we consider instantaneous for the simulation purposes).

Fig. 2(c) shows the time evolution of the simulated resistance of the sample  during the application of the  electroforming protocol, $V_s(t)$, shown in Fig. 2(b).  
The agreement between  the simulated and the experimental curve is remarkable.
After a transient, in  which the device resistance fluctuates
for a time scale of the order of 100 s,  the resistance finally stabilizes in  the remanent resistance vs. voltage loop shown in Fig. 2(d). Notice that the simulations perfectly capture the time scale of this process with an attained stable resistance loop that reproduces most of the characteristics of the experimental one.

Fig. 2(c) additionally shows the time  evolution of the relative fraction (percentage)  of OV remaining in the sample $\delta_0\equiv N_{VO}(t)/N_{V0}(0^+) \%$  (that is the percentage ratio between the total number of OV at time t  and at the initial time). A saturation  close to 75$\%$, once the stable resistant loop is attained, is clearly observed.
This might correspond to an equilibrium state between the device and the environment with no subsequent OV exchange during the stable memristive cycling, as we address in the next section. 

\section{\label{sec:2}Stable memristive behavior: experiments and simulations}

We have measured and simulated the stable memristive response of another device with a larger area ($\approx 1.2x10^4$ $\mu$m$^2$) than the previous one, which was electroformed in a similar way than described before and stimulated with different writing voltage cycles, characterized by their maximum (minimum) excursions V$_{MAX}>0$ (-V$_{MIN}>0$). Fig. 3(a) displays the case of V$_{MAX}$ < -V$_{MIN}$, with a remanent resistance loop vs.  voltage that presents a clockwise (CW) evolution, while Fig. 3(b) displays the case of V$_{MAX}$ > -V$_{MIN}$, characterized by a loop with a counter-clockwise (CCW) evolution.

\begin{figure} 
\centering
\includegraphics[width=0.95\linewidth]{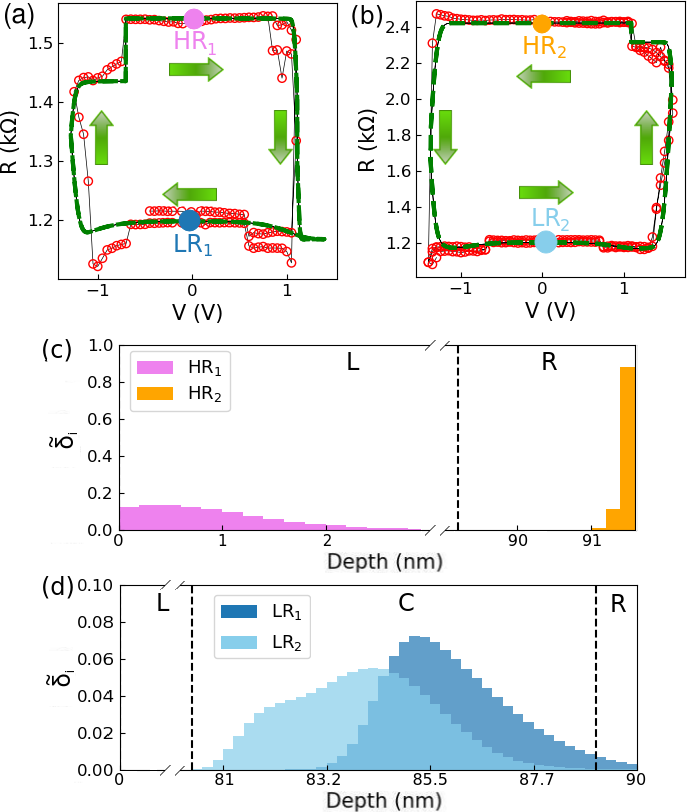}
\caption{(a) CW remanent resistance vs. voltage loops for a $\approx 1.2x10^4$ $\mu$m$^2$ Pt/TiO$_x$/TaO$_y$/Pt device. Experimental resistance values are shown in red and simulated ones are shown in green; (b) CCW loops for the same device. The transition between CW and CCW loops is attained by tuning the maximum and minimum writing voltage excursions; (c)  OV profiles  correspondent to the HR$_1$ state of CW (pink) and HR$_2$ of CCW (orange) simulated loops; (d) OV profiles correspondent to the LR$_1$ state of CW (dark blue) and  LR$_2$ of CCW (light blue) simulated  loops. The OV density per site ($\widetilde{\delta}_i)$ is normalized in  terms of  $\delta = 0.0039$, the initial (uniform) OV density considered.}
\label{Figura2}
\end{figure}


We notice that both remanent resistance vs. voltage loops share quite the same low resistance value (LR$_{1}$ $\sim$ LR$_{2} \approx$ 1.2 k$\Omega$) but they differ in their high resistance state ($HR_1 \approx$ 1.5 k$\Omega$ for the CW loop and $HR_2  \approx$ 2.4 k$\Omega$ for the CCW loop). We  remark that the resistance levels of the CW loop (Fig. 3(a)) are around one order of magnitude lower than those found for the device described previously, with a similar evolution (recall Fig. 2(d)). This indicates that the resistance levels increase as the device area is decreased, consistently with a non-filamentary, area distributed memristive effect, as previously reported for other simple oxides-based memristive systems \cite{hsu_2014,wang_2015,lee_2018,ferreyra_2020}. 

In addition, our results shows the possibility of tuning the evolution (circulation) of the remanent resistance loop in a reversible way, as it  was previously found in Pt/TaO$_y$/TaO$_h$/Pt devices and  linked to the  control  at the nanoscale of the  OV dynamics through asymmetric electrical stimuli, allowing the selective activation/deactivation of both oxide/Pt interfaces \cite{ferreyra_2020}. 

In Figs. 3(a) and 3(b) are shown the simulations of both CW and CCW resistance loops (dashed lines), displaying an excellent agreement with the experimental ones (dotted lines). 
The distinctive feature of these simulations is that they are performed  conserving the total amount of OV present in the system. In other words, our results confirm that the interaction between the device and the environment is restricted to the electroforming process, but later on the memristive effect relies exclusively on the internal redistribution of OV between different zones of the device. This result is at odds with Ref. \cite{kim_2016}, where it was suggested that oxygen exchange takes place during the stable memristive cycling for ohmic top metal/oxide interfaces. The difference in our case might rely on the existence of an energy barrier (i.e. Schottky type) at the top Pt/TiO$_x$ interface \cite{ma_2017}. The OV distribution along the device is shown for the HR$_1$, HR$_2$ and LR$_1 \sim$ LR$_2$ states, pointed out in the correspondent resistance loops displayed in Fig. 3.

For the   low resistance states, in both CW and CCW loops, OV accumulate at the central Ti and Ta intermixing (C) zone, reducing its resistance and driving the overall drop of the total device resistance. 
Notice that the residual penetration of OV in the R zone in the case of the CW loop is responsible for the tiny difference between the LR$_1$ and LR$_2$ values. 
Additionally for the CW(CCW) loop the HR$_1$(HR$_2$) state  corresponds to OV located mainly at the L(R) interface, while in both cases the C region, being depleted from OV, is responsible for  the (high) resistance value of the device.


We notice that in order to properly simulate the experimental electrical behavior, it should be assumed that the C zone has the highest residual resistivity of the device. This implies that the RS effect is dominated by the resistivity changes of zone C, driven by OV electromigration between this zone and the two metal/oxide interfaces, depending on the polarity and the asymmetry of the applied stimulus.


Starting the simulations from an  OV distribution compatible with the post forming HR$_1$ state, the SET  transition to the LR$_1$ state in the CW loop takes place when the  OV, initially located at  the L interface,  have been driven  to  the C zone under the positive
SET stimulus. For the CCW loop,   the  LR$_2$ state is attained  
when the OV,  former located at the R interface, are drifted to the C zone under  the 
negative SET voltage. 

The latter analysis can be complemented by fitting the dynamic current-voltage (I-V) curves, recorded simultaneously with the CW and CCW remanent resistance loops (see Ref.\cite{ferreyra_2020} for further experimental details), after proposing an equivalent circuit. Figs. 4(a) and (b) display the dynamic I-V curves related to the  remanent resistance loops with CW and CCW circulations, respectively. It is seen that the I-V curves display an inverse circulation in relation to the remanent resistance loops, as expected. They show a non-linear evolution indicating the presence of non-ohmic transport mechanisms, as usually found in capacitor-like structures with memristive non-filamentary behavior \cite{marlasca_2013,Acevedo_2017,ferreyra_2020,acevedo_2020}. To perform the fittings we considered the $\gamma$ = dLn(I)/dLn(V) parameter representation, firstly introduced in Ref. \cite{acha_2017} and which was proved as a suitable way for undisclosing multiple conduction mechanisms,  usually found in oxide based memristors \cite{Acevedo_2017,ferreyra_2020,acevedo_2020}.

\begin{figure} 
\centering
\includegraphics[width=0.95\linewidth]{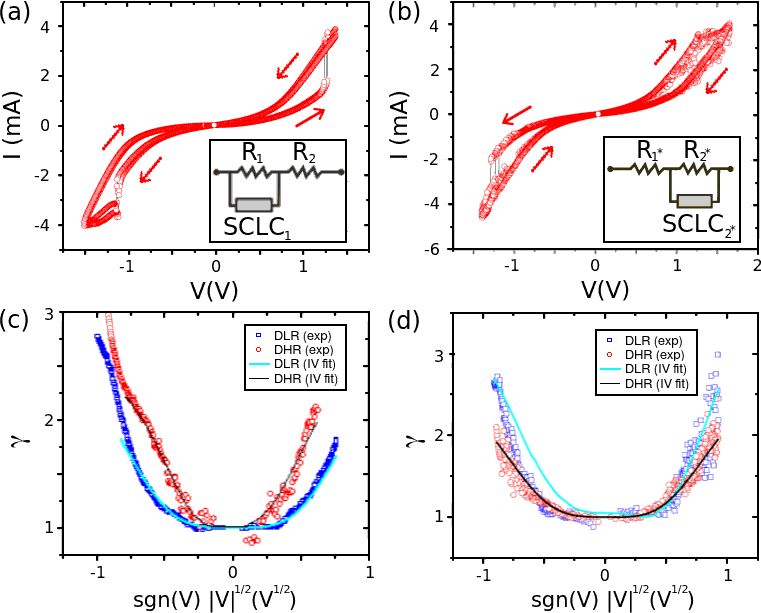}
\caption{Dynamic I-V curves recorded simultaneously with CW (a) and CCW (b) remanent resistance loops. The different I-V circulations -mirrored with respect to the remanent resistance loops- are attained by changing the maximum and minimum voltage excursions (see text for details). The insets show the minimal equivalent circuits necessary to fit the experimental data; (c), (d) $\gamma$ vs. $V^{1/2}$ representation of the Dynamic Low and High Resistance states (DHR and DLR, respectively), obtained from the I-V curves displayed in (a) and (b). Symbols correspond to the experimental data, while the fittings (see the text for details) are shown in solid lines.}
\label{Figura4}
\end{figure}

Figs. 4(c) and (d) display the $\gamma$ vs. V$^{1/2}$ behavior derived from the aforementioned I-V curves, both for low and high resistance branches, which we name as Dynamic Low (High) Resistance state or DLR (DHR). We notice that only voltages with absolute values lower that the SET/RESET ones are considered. First, we stress the existence of ohmic conduction for low voltages ($\gamma$ $\approx$ 1); for higher voltages, a non-linear conduction mechanism, which we identify as Space Charge Limited Current (SCLC) conduction with traps, prevails, as $\gamma$ increases smoothly with V and reaches values $\geq$ 2
\cite{Murga_1970}. The simplest circuit representations consistent with the evolution of the $\gamma$ parameter with voltage are displayed as insets in Figs. 4(a) and 4(b). For the case of the I-V displayed in Fig. 4(a) -corresponding to a CW remanent resistance loop-, the equivalent circuit comprises a parallel combination of a resistor $R_{1}$ and a SCLC channel ($SCLC_{1}$), in series with a resistor $R_{2}$ (see the inset of Fig. 4(a)). The current flowing through the device can be described by I = $I_{R_{1}}$ + I$_{SCLC_{1}}$ = (V - I$R_{2}$) / R$_{1}$ +  $A_{1}$ (V-I$R_{2}$)$^{n_{1}}$, where $I_{R_{1}}$ and $I_{SCLC_{1}}$ are the currents through the $R_{1}$ resistor and the SCLC element, respectively. $A_{1}$ is related to the mobility, the dielectric constant and the width of the transport channel and $n_{1}$ is an exponent $\geq$ 2. This implicit equation was solved numerically in order to fit the experimental $\gamma$ vs. V$^{1/2}$ curves, both for DLR and DHR, by determining in each case the fitting parameters R$_{1}$, A$_{1}$, $n_{1}$ and R$_{2}$. The results of the fittings can be observed in Fig. 4(c), showing a very good agreement with the experimental data \footnote{We also tested the possible contributions of Schottky diodes present at both metal-oxide interfaces, finding that they don't significantly contribute to the electronic transport in the range of (small) voltages used to perform the fittings. They would only contribute with a small part of the conduction in the range of higher voltages, where some deviations between the fits and the experimental values can be observed.}. It is found that the transition from DLR to DHR is driven by changes in the non-linear SCLC$_{1}$ element (A$_{1}$ and $n_{1}$ are $\approx$ 64\% and $\approx$ 17\%  lower for the DHR state, respectively) and its parallel leakage channel R$_{1}$. A similar analysis can be made for the case of the I-V curve displayed in Fig. 4(b) -corresponding to a CCW remanent resistance loop-, where the equivalent circuit in this case corresponds to a resistor R$_{1}$ in series with the parallel combination of a SCLC element (SCLC$_{2}$) and a resistor R$_{2}$ (see the inset of Fig. 4(b)). The corresponding $\gamma$ vs. V$^{1/2}$ fittings for both DLW and DHR states are shown in Fig. 4(d), again with a good agreement between the experimental and calculated curves. In this case, the memristive effect is dominated by changes in the element SCLC$_{2}$ (A$_{2}$ and $n_{2}$ are $\approx$  25\% and $\approx$ 7\%  lower for the DHR state, respectively) and its leakage channel R$_{2}$. From the analysis of the two presented cases (I-V curves with opposite circulations) it is found that both the transport mechanism and memristive effect are strongly dependant on a SCLC channel. We recall that SCLC is a bulk conduction mechanism, which supports our previous statement that the resistance change is not dominated by the oxide/metal interfaces but by a bulk zone of the device in between both metal/oxide interfaces, including the intermixed TiO$_x$/TaO$_y$ interface.

\section{Concluding remarks}

From a combination of electrical measurements, analytic characterization and  modelling, we have unveiled  the role of OV exchange between  Pt/TiO$_x$/TaO$_y$/Pt memristive devices and the environment. Our microscopy experiments show a clear oxidation process of both TiO$_x$ and TaO$_y$ layers during the electroforming process, validated by our numerical simulations based on the VEOV model. It is reasonable to assume that the top Pt electrode, with a microstructure of columns and grain boundaries, behaves as a permeable layer that allows oxygen transport between the environment and the oxide bilayer trough grain boundaries, as has been previously suggested in Ref. \cite{zurhelle_2022}. In addition to this oxidation process, electroforming also shows the formation of oxide nanograins -likely at the first stages of the forming process-, reflecting the presence of strong thermal effects that trigger the partial crystallization of the oxide bilayer. Once the electroforming process is complete, the system is able to switch between stable low and high resistance states in two different ways: if the writing voltages are such that V$_{MAX}$ > -V$_{MIN}$ the remanent resistance shows a CCW evolution, while if V$_{MAX}$ < -V$_{MIN}$ the loops display a CW evolution. For both cases, the remanent resistance loops can be numerically reproduced by assuming that the system  
maintains a constant number of OV, indicating that it is in equilibrium with the environment. 
This difference with respect to the electroforming process can be related to the fact that V$_{MAX}$ and -V$_{MIN}$ are lower that the forming maximum voltage -V$_{FO}$. However, other features such as the nature of the top metal/oxide interface might also play a significant role \cite{kim_2016}. Based on our numerical simulations, it was established that the OV dynamics for the stable CW (CCW) loop is constrained to the OV exchange between TiO$_x$ (TaO$_y$) layer and the central device zone comprising the TiO$_x$/TaO$_y$ interface, where Ti and Ta interdiffusion was observed. This central zone was also shown to dominate the electrical transport and to control the resistive changes of the device for both cases. Our work provides relevant information for the design of reliable binary oxides memristive systems, which are strong candidates for the implementation of neuromorphic computing devices such as physical neural networks \cite{prez15}.

\section*{Acknowlegments}
 
We acknowledge support from UNCuyo (06/C591), ANPCyT (PICT2017-1836, PICT2019-02781, PICT2019-0654 and PICT2020A-00415) and EU-H2020-RISE project "MELON" (Grant No. 872631). We also acknowledge the LMA-Universidad de Zaragoza for offering access to the microscopy instruments.
MJS acknowledges the hospitality of the LPMC, Université of Picardie Jules Verne.

\section*{Data Availability Statement}

The data that support the findings of
this study are available from the
corresponding author upon reasonable
request

\bibliography{references}

\begin{thebibliography}{53}%
\makeatletter
\providecommand \@ifxundefined [1]{%
 \@ifx{#1\undefined}
}%
\providecommand \@ifnum [1]{%
 \ifnum #1\expandafter \@firstoftwo
 \else \expandafter \@secondoftwo
 \fi
}%
\providecommand \@ifx [1]{%
 \ifx #1\expandafter \@firstoftwo
 \else \expandafter \@secondoftwo
 \fi
}%
\providecommand \natexlab [1]{#1}%
\providecommand \enquote  [1]{``#1''}%
\providecommand \bibnamefont  [1]{#1}%
\providecommand \bibfnamefont [1]{#1}%
\providecommand \citenamefont [1]{#1}%
\providecommand \href@noop [0]{\@secondoftwo}%
\providecommand \href [0]{\begingroup \@sanitize@url \@href}%
\providecommand \@href[1]{\@@startlink{#1}\@@href}%
\providecommand \@@href[1]{\endgroup#1\@@endlink}%
\providecommand \@sanitize@url [0]{\catcode `\\12\catcode `\$12\catcode
  `\&12\catcode `\#12\catcode `\^12\catcode `\_12\catcode `\%12\relax}%
\providecommand \@@startlink[1]{}%
\providecommand \@@endlink[0]{}%
\providecommand \url  [0]{\begingroup\@sanitize@url \@url }%
\providecommand \@url [1]{\endgroup\@href {#1}{\urlprefix }}%
\providecommand \urlprefix  [0]{URL }%
\providecommand \Eprint [0]{\href }%
\providecommand \doibase [0]{http://dx.doi.org/}%
\providecommand \selectlanguage [0]{\@gobble}%
\providecommand \bibinfo  [0]{\@secondoftwo}%
\providecommand \bibfield  [0]{\@secondoftwo}%
\providecommand \translation [1]{[#1]}%
\providecommand \BibitemOpen [0]{}%
\providecommand \bibitemStop [0]{}%
\providecommand \bibitemNoStop [0]{.\EOS\space}%
\providecommand \EOS [0]{\spacefactor3000\relax}%
\providecommand \BibitemShut  [1]{\csname bibitem#1\endcsname}%
\let\auto@bib@innerbib\@empty
\bibitem [{\citenamefont {Sawa}(2008)}]{saw_2008}%
  \BibitemOpen
  \bibfield  {author} {\bibinfo {author} {\bibfnamefont {A.}~\bibnamefont
  {Sawa}},\ }\href@noop {} {\bibfield  {journal} {\bibinfo  {journal} {Mater.
  Today}\ }\textbf {\bibinfo {volume} {11}},\ \bibinfo {pages} {28} (\bibinfo
  {year} {2008})}\BibitemShut {NoStop}%
\bibitem [{\citenamefont {Ielmini}\ and\ \citenamefont
  {Waser}(2016)}]{iel_2016}%
  \BibitemOpen
  \bibfield  {author} {\bibinfo {author} {\bibfnamefont {D.}~\bibnamefont
  {Ielmini}}\ and\ \bibinfo {author} {\bibfnamefont {R.}~\bibnamefont
  {Waser}},\ }\href@noop {} {\emph {\bibinfo {title} {Resistive Switching: From
  Fundamentals of Nanoionic Redox Processes to Memristive Device
  Applications}}}\ (\bibinfo  {publisher} {Weinheim: Wiley-VCH},\ \bibinfo
  {year} {2016})\BibitemShut {NoStop}%
\bibitem [{\citenamefont {Yu}(2017)}]{yu_2017}%
  \BibitemOpen
  \bibfield  {author} {\bibinfo {author} {\bibfnamefont {S.}~\bibnamefont
  {Yu}},\ }\href@noop {} {\emph {\bibinfo {title} {Neuro-Inspiring Computing
  Using Resistive Synaptic Devices}}}\ (\bibinfo  {publisher} {Cham:
  Springer},\ \bibinfo {year} {2017})\BibitemShut {NoStop}%
\bibitem [{\citenamefont {Traversa}\ and\ \citenamefont
  {Di~Ventra}(2015)}]{trave_2015}%
  \BibitemOpen
  \bibfield  {author} {\bibinfo {author} {\bibfnamefont {F.~L.}\ \bibnamefont
  {Traversa}}\ and\ \bibinfo {author} {\bibfnamefont {M.}~\bibnamefont
  {Di~Ventra}},\ }\href@noop {} {\bibfield  {journal} {\bibinfo  {journal}
  {IEEE transactions on neural networks and learning systems}\ }\textbf
  {\bibinfo {volume} {26}},\ \bibinfo {pages} {2702} (\bibinfo {year}
  {2015})}\BibitemShut {NoStop}%
\bibitem [{\citenamefont {Yang}\ \emph {et~al.}(2008)\citenamefont {Yang},
  \citenamefont {Pickett}, \citenamefont {Li}, \citenamefont {Ohlberg},
  \citenamefont {Stewart},\ and\ \citenamefont {Williams}}]{yang_2008}%
  \BibitemOpen
  \bibfield  {author} {\bibinfo {author} {\bibfnamefont {J.~J.}\ \bibnamefont
  {Yang}}, \bibinfo {author} {\bibfnamefont {M.~D.}\ \bibnamefont {Pickett}},
  \bibinfo {author} {\bibfnamefont {X.}~\bibnamefont {Li}}, \bibinfo {author}
  {\bibfnamefont {D.~A.~A.}\ \bibnamefont {Ohlberg}}, \bibinfo {author}
  {\bibfnamefont {D.~R.}\ \bibnamefont {Stewart}}, \ and\ \bibinfo {author}
  {\bibfnamefont {R.~S.}\ \bibnamefont {Williams}},\ }\href@noop {} {\bibfield
  {journal} {\bibinfo  {journal} {Nature Nanotechnology}\ }\textbf {\bibinfo
  {volume} {3}},\ \bibinfo {pages} {429} (\bibinfo {year} {2008})}\BibitemShut
  {NoStop}%
\bibitem [{\citenamefont {Gunkel}\ \emph {et~al.}(2020)\citenamefont {Gunkel},
  \citenamefont {Christensen}, \citenamefont {Chen},\ and\ \citenamefont
  {Pryds}}]{gun20}%
  \BibitemOpen
  \bibfield  {author} {\bibinfo {author} {\bibfnamefont {F.}~\bibnamefont
  {Gunkel}}, \bibinfo {author} {\bibfnamefont {D.~V.}\ \bibnamefont
  {Christensen}}, \bibinfo {author} {\bibfnamefont {Y.~Z.}\ \bibnamefont
  {Chen}}, \ and\ \bibinfo {author} {\bibfnamefont {N.}~\bibnamefont {Pryds}},\
  }\href@noop {} {\bibfield  {journal} {\bibinfo  {journal} {Appl. Phys.
  Lett.}\ }\textbf {\bibinfo {volume} {116}},\ \bibinfo {pages} {120505}
  (\bibinfo {year} {2020})}\BibitemShut {NoStop}%
\bibitem [{\citenamefont {Rubi}\ \emph {et~al.}(2013)\citenamefont {Rubi},
  \citenamefont {Tesler}, \citenamefont {Alposta}, \citenamefont {Kalstein},
  \citenamefont {Ghenzi}, \citenamefont {Gomez-Marlasca}, \citenamefont
  {Rozenberg},\ and\ \citenamefont {Levy}}]{rubi_2013}%
  \BibitemOpen
  \bibfield  {author} {\bibinfo {author} {\bibfnamefont {D.}~\bibnamefont
  {Rubi}}, \bibinfo {author} {\bibfnamefont {F.}~\bibnamefont {Tesler}},
  \bibinfo {author} {\bibfnamefont {I.}~\bibnamefont {Alposta}}, \bibinfo
  {author} {\bibfnamefont {A.}~\bibnamefont {Kalstein}}, \bibinfo {author}
  {\bibfnamefont {N.}~\bibnamefont {Ghenzi}}, \bibinfo {author} {\bibfnamefont
  {F.}~\bibnamefont {Gomez-Marlasca}}, \bibinfo {author} {\bibfnamefont
  {M.}~\bibnamefont {Rozenberg}}, \ and\ \bibinfo {author} {\bibfnamefont
  {P.}~\bibnamefont {Levy}},\ }\href {\doibase 10.1063/1.4826484} {\bibfield
  {journal} {\bibinfo  {journal} {Applied Physics Letters}\ }\textbf {\bibinfo
  {volume} {103}},\ \bibinfo {pages} {163506} (\bibinfo {year}
  {2013})}\BibitemShut {NoStop}%
\bibitem [{\citenamefont {Muenstermann}\ \emph {et~al.}(2010)\citenamefont
  {Muenstermann}, \citenamefont {Menke}, \citenamefont {Dittmann},\ and\
  \citenamefont {Waser}}]{muenster_2010}%
  \BibitemOpen
  \bibfield  {author} {\bibinfo {author} {\bibfnamefont {R.}~\bibnamefont
  {Muenstermann}}, \bibinfo {author} {\bibfnamefont {T.}~\bibnamefont {Menke}},
  \bibinfo {author} {\bibfnamefont {R.}~\bibnamefont {Dittmann}}, \ and\
  \bibinfo {author} {\bibfnamefont {R.}~\bibnamefont {Waser}},\ }\href
  {\doibase https://doi.org/10.1002/adma.201001872} {\bibfield  {journal}
  {\bibinfo  {journal} {Advanced Materials}\ }\textbf {\bibinfo {volume}
  {22}},\ \bibinfo {pages} {4819} (\bibinfo {year} {2010})},\ \Eprint
  {http://arxiv.org/abs/https://onlinelibrary.wiley.com/doi/pdf/10.1002/adma.201001872}
  {https://onlinelibrary.wiley.com/doi/pdf/10.1002/adma.201001872} \BibitemShut
  {NoStop}%
\bibitem [{\citenamefont {Hwan~Kim}\ \emph {et~al.}(2011)\citenamefont
  {Hwan~Kim}, \citenamefont {Ho~Lee}, \citenamefont {Yeong~Seok}, \citenamefont
  {Ji~Song}, \citenamefont {Ho~Yoon}, \citenamefont {Jean~Yoon}, \citenamefont
  {Hwan~Lee}, \citenamefont {Min~Kim}, \citenamefont {Dong~Lee}, \citenamefont
  {Wook~Ryu}, \citenamefont {Joo~Park},\ and\ \citenamefont
  {Seong~Hwang}}]{ryu_2011}%
  \BibitemOpen
  \bibfield  {author} {\bibinfo {author} {\bibfnamefont {G.}~\bibnamefont
  {Hwan~Kim}}, \bibinfo {author} {\bibfnamefont {J.}~\bibnamefont {Ho~Lee}},
  \bibinfo {author} {\bibfnamefont {J.}~\bibnamefont {Yeong~Seok}}, \bibinfo
  {author} {\bibfnamefont {S.}~\bibnamefont {Ji~Song}}, \bibinfo {author}
  {\bibfnamefont {J.}~\bibnamefont {Ho~Yoon}}, \bibinfo {author} {\bibfnamefont
  {K.}~\bibnamefont {Jean~Yoon}}, \bibinfo {author} {\bibfnamefont
  {M.}~\bibnamefont {Hwan~Lee}}, \bibinfo {author} {\bibfnamefont
  {K.}~\bibnamefont {Min~Kim}}, \bibinfo {author} {\bibfnamefont
  {H.}~\bibnamefont {Dong~Lee}}, \bibinfo {author} {\bibfnamefont
  {S.}~\bibnamefont {Wook~Ryu}}, \bibinfo {author} {\bibfnamefont
  {T.}~\bibnamefont {Joo~Park}}, \ and\ \bibinfo {author} {\bibfnamefont
  {C.}~\bibnamefont {Seong~Hwang}},\ }\href {\doibase 10.1063/1.3600784}
  {\bibfield  {journal} {\bibinfo  {journal} {Appl. Phys. Lett.}\ }\textbf
  {\bibinfo {volume} {98}},\ \bibinfo {pages} {262901} (\bibinfo {year}
  {2011})}\BibitemShut {NoStop}%
\bibitem [{\citenamefont {Wang}\ \emph {et~al.}(2015)\citenamefont {Wang},
  \citenamefont {Lin}, \citenamefont {Wang}, \citenamefont {Lin},\ and\
  \citenamefont {Hou}}]{wang_2015}%
  \BibitemOpen
  \bibfield  {author} {\bibinfo {author} {\bibfnamefont {Y.-F.}\ \bibnamefont
  {Wang}}, \bibinfo {author} {\bibfnamefont {Y.-C.}\ \bibnamefont {Lin}},
  \bibinfo {author} {\bibfnamefont {I.-T.}\ \bibnamefont {Wang}}, \bibinfo
  {author} {\bibfnamefont {T.-P.}\ \bibnamefont {Lin}}, \ and\ \bibinfo
  {author} {\bibfnamefont {T.-H.}\ \bibnamefont {Hou}},\ }\href {\doibase
  10.1038/srep10150} {\bibfield  {journal} {\bibinfo  {journal} {Sci. Rep.}\
  }\textbf {\bibinfo {volume} {5}},\ \bibinfo {pages} {10150} (\bibinfo {year}
  {2015})}\BibitemShut {NoStop}%
\bibitem [{\citenamefont {Choi}\ \emph {et~al.}(2018)\citenamefont {Choi},
  \citenamefont {Jang}, \citenamefont {Moon}, \citenamefont {Kim},
  \citenamefont {Jeong}, \citenamefont {Jang}, \citenamefont {Lee},\ and\
  \citenamefont {Wang}}]{choi_2018}%
  \BibitemOpen
  \bibfield  {author} {\bibinfo {author} {\bibfnamefont {S.}~\bibnamefont
  {Choi}}, \bibinfo {author} {\bibfnamefont {S.}~\bibnamefont {Jang}}, \bibinfo
  {author} {\bibfnamefont {J.-H.}\ \bibnamefont {Moon}}, \bibinfo {author}
  {\bibfnamefont {J.~C.}\ \bibnamefont {Kim}}, \bibinfo {author} {\bibfnamefont
  {H.~Y.}\ \bibnamefont {Jeong}}, \bibinfo {author} {\bibfnamefont
  {P.}~\bibnamefont {Jang}}, \bibinfo {author} {\bibfnamefont {K.-J.}\
  \bibnamefont {Lee}}, \ and\ \bibinfo {author} {\bibfnamefont
  {G.}~\bibnamefont {Wang}},\ }\href {\doibase 10.1038/s41427-018-0101-y}
  {\bibfield  {journal} {\bibinfo  {journal} {NPG Asia Mater.}\ }\textbf
  {\bibinfo {volume} {10}},\ \bibinfo {pages} {1097 } (\bibinfo {year}
  {2018})}\BibitemShut {NoStop}%
\bibitem [{\citenamefont {Yang}\ \emph {et~al.}(2010)\citenamefont {Yang},
  \citenamefont {Zhang}, \citenamefont {Strachan}, \citenamefont {Miao},
  \citenamefont {Pickett}, \citenamefont {Kelley}, \citenamefont
  {Medeiros-Ribeiro},\ and\ \citenamefont {Williams}}]{yang_2010}%
  \BibitemOpen
  \bibfield  {author} {\bibinfo {author} {\bibfnamefont {J.~J.}\ \bibnamefont
  {Yang}}, \bibinfo {author} {\bibfnamefont {M.-X.}\ \bibnamefont {Zhang}},
  \bibinfo {author} {\bibfnamefont {J.~P.}\ \bibnamefont {Strachan}}, \bibinfo
  {author} {\bibfnamefont {F.}~\bibnamefont {Miao}}, \bibinfo {author}
  {\bibfnamefont {M.~D.}\ \bibnamefont {Pickett}}, \bibinfo {author}
  {\bibfnamefont {R.~D.}\ \bibnamefont {Kelley}}, \bibinfo {author}
  {\bibfnamefont {G.}~\bibnamefont {Medeiros-Ribeiro}}, \ and\ \bibinfo
  {author} {\bibfnamefont {R.~S.}\ \bibnamefont {Williams}},\ }\href {\doibase
  10.1063/1.3524521} {\bibfield  {journal} {\bibinfo  {journal} {Appl. Phys.
  Lett.}\ }\textbf {\bibinfo {volume} {97}},\ \bibinfo {pages} {232102}
  (\bibinfo {year} {2010})}\BibitemShut {NoStop}%
\bibitem [{\citenamefont {Lee}\ \emph {et~al.}(2018)\citenamefont {Lee},
  \citenamefont {Park}, \citenamefont {Seo}, \citenamefont {Kwon},
  \citenamefont {Lee}, \citenamefont {Kim}, \citenamefont {Jung}, \citenamefont
  {You}, \citenamefont {Lee}, \citenamefont {Kim}, \citenamefont {Pang},
  \citenamefont {Seo}, \citenamefont {Hwang},\ and\ \citenamefont
  {Park}}]{lee_2018}%
  \BibitemOpen
  \bibfield  {author} {\bibinfo {author} {\bibfnamefont {M.-J.}\ \bibnamefont
  {Lee}}, \bibinfo {author} {\bibfnamefont {G.-S.}\ \bibnamefont {Park}},
  \bibinfo {author} {\bibfnamefont {D.~H.}\ \bibnamefont {Seo}}, \bibinfo
  {author} {\bibfnamefont {S.~M.}\ \bibnamefont {Kwon}}, \bibinfo {author}
  {\bibfnamefont {H.-J.}\ \bibnamefont {Lee}}, \bibinfo {author} {\bibfnamefont
  {J.-S.}\ \bibnamefont {Kim}}, \bibinfo {author} {\bibfnamefont
  {M.}~\bibnamefont {Jung}}, \bibinfo {author} {\bibfnamefont {C.-Y.}\
  \bibnamefont {You}}, \bibinfo {author} {\bibfnamefont {H.}~\bibnamefont
  {Lee}}, \bibinfo {author} {\bibfnamefont {H.-G.}\ \bibnamefont {Kim}},
  \bibinfo {author} {\bibfnamefont {S.-B.}\ \bibnamefont {Pang}}, \bibinfo
  {author} {\bibfnamefont {S.}~\bibnamefont {Seo}}, \bibinfo {author}
  {\bibfnamefont {H.}~\bibnamefont {Hwang}}, \ and\ \bibinfo {author}
  {\bibfnamefont {S.~K.}\ \bibnamefont {Park}},\ }\href {\doibase
  10.1021/acsami.8b09046} {\bibfield  {journal} {\bibinfo  {journal} {ACS Appl.
  Mater. Interfaces}\ }\textbf {\bibinfo {volume} {10}},\ \bibinfo {pages}
  {29757 } (\bibinfo {year} {2018})}\BibitemShut {NoStop}%
\bibitem [{\citenamefont {Böttger}\ \emph {et~al.}(2020)\citenamefont
  {Böttger}, \citenamefont {von Witzleben}, \citenamefont {Havel},
  \citenamefont {Fleck}, \citenamefont {Rana}, \citenamefont {Waser},\ and\
  \citenamefont {Menzel}}]{bottger_2020}%
  \BibitemOpen
  \bibfield  {author} {\bibinfo {author} {\bibfnamefont {U.}~\bibnamefont
  {Böttger}}, \bibinfo {author} {\bibfnamefont {M.}~\bibnamefont {von
  Witzleben}}, \bibinfo {author} {\bibfnamefont {V.}~\bibnamefont {Havel}},
  \bibinfo {author} {\bibfnamefont {K.}~\bibnamefont {Fleck}}, \bibinfo
  {author} {\bibfnamefont {V.}~\bibnamefont {Rana}}, \bibinfo {author}
  {\bibfnamefont {R.}~\bibnamefont {Waser}}, \ and\ \bibinfo {author}
  {\bibfnamefont {S.}~\bibnamefont {Menzel}},\ }\href {\doibase
  10.1038/s41598-020-73254-2} {\bibfield  {journal} {\bibinfo  {journal} {Sci.
  Rep.}\ }\textbf {\bibinfo {volume} {10}},\ \bibinfo {pages} {16391} (\bibinfo
  {year} {2020})}\BibitemShut {NoStop}%
\bibitem [{\citenamefont {Shi}\ \emph {et~al.}(2017)\citenamefont {Shi},
  \citenamefont {Xu}, \citenamefont {Wang}, \citenamefont {Zhao}, \citenamefont
  {Liu}, \citenamefont {Ma},\ and\ \citenamefont {Liu}}]{shi_2017}%
  \BibitemOpen
  \bibfield  {author} {\bibinfo {author} {\bibfnamefont {K.~X.}\ \bibnamefont
  {Shi}}, \bibinfo {author} {\bibfnamefont {H.~Y.}\ \bibnamefont {Xu}},
  \bibinfo {author} {\bibfnamefont {Z.~Q.}\ \bibnamefont {Wang}}, \bibinfo
  {author} {\bibfnamefont {X.~N.}\ \bibnamefont {Zhao}}, \bibinfo {author}
  {\bibfnamefont {W.~Z.}\ \bibnamefont {Liu}}, \bibinfo {author} {\bibfnamefont
  {J.~G.}\ \bibnamefont {Ma}}, \ and\ \bibinfo {author} {\bibfnamefont {Y.~C.}\
  \bibnamefont {Liu}},\ }\href {\doibase 10.1063/1.5002571} {\bibfield
  {journal} {\bibinfo  {journal} {Appl. Phys. Lett.}\ }\textbf {\bibinfo
  {volume} {111}},\ \bibinfo {pages} {223505} (\bibinfo {year}
  {2017})}\BibitemShut {NoStop}%
\bibitem [{\citenamefont {Wedig}\ \emph {et~al.}(2016)\citenamefont {Wedig},
  \citenamefont {Luebben}, \citenamefont {Cho}, \citenamefont {Moors},
  \citenamefont {Skaja}, \citenamefont {Rana}, \citenamefont {Hasegawa},
  \citenamefont {Adepalli}, \citenamefont {Yildiz}, \citenamefont {Waser},\
  and\ \citenamefont {Valov}}]{wedig_2015}%
  \BibitemOpen
  \bibfield  {author} {\bibinfo {author} {\bibfnamefont {A.}~\bibnamefont
  {Wedig}}, \bibinfo {author} {\bibfnamefont {M.}~\bibnamefont {Luebben}},
  \bibinfo {author} {\bibfnamefont {D.-Y.}\ \bibnamefont {Cho}}, \bibinfo
  {author} {\bibfnamefont {M.}~\bibnamefont {Moors}}, \bibinfo {author}
  {\bibfnamefont {K.}~\bibnamefont {Skaja}}, \bibinfo {author} {\bibfnamefont
  {V.}~\bibnamefont {Rana}}, \bibinfo {author} {\bibfnamefont {T.}~\bibnamefont
  {Hasegawa}}, \bibinfo {author} {\bibfnamefont {K.~K.}\ \bibnamefont
  {Adepalli}}, \bibinfo {author} {\bibfnamefont {B.}~\bibnamefont {Yildiz}},
  \bibinfo {author} {\bibfnamefont {R.}~\bibnamefont {Waser}}, \ and\ \bibinfo
  {author} {\bibfnamefont {I.}~\bibnamefont {Valov}},\ }\href {\doibase
  10.1038/nnano.2015.221} {\bibfield  {journal} {\bibinfo  {journal} {Nat.
  Nanotechnol.}\ }\textbf {\bibinfo {volume} {11}},\ \bibinfo {pages} {67 }
  (\bibinfo {year} {2016})}\BibitemShut {NoStop}%
\bibitem [{\citenamefont {Park}\ \emph {et~al.}(2015)\citenamefont {Park},
  \citenamefont {Song}, \citenamefont {Kim}, \citenamefont {Kim}, \citenamefont
  {Chung}, \citenamefont {Kim}, \citenamefont {Lee}, \citenamefont {Kim},
  \citenamefont {Choi},\ and\ \citenamefont {Hwang}}]{park_2015}%
  \BibitemOpen
  \bibfield  {author} {\bibinfo {author} {\bibfnamefont {T.~H.}\ \bibnamefont
  {Park}}, \bibinfo {author} {\bibfnamefont {S.~J.}\ \bibnamefont {Song}},
  \bibinfo {author} {\bibfnamefont {H.~J.}\ \bibnamefont {Kim}}, \bibinfo
  {author} {\bibfnamefont {S.~G.}\ \bibnamefont {Kim}}, \bibinfo {author}
  {\bibfnamefont {S.}~\bibnamefont {Chung}}, \bibinfo {author} {\bibfnamefont
  {B.~Y.}\ \bibnamefont {Kim}}, \bibinfo {author} {\bibfnamefont {K.~J.}\
  \bibnamefont {Lee}}, \bibinfo {author} {\bibfnamefont {K.~M.}\ \bibnamefont
  {Kim}}, \bibinfo {author} {\bibfnamefont {B.~J.}\ \bibnamefont {Choi}}, \
  and\ \bibinfo {author} {\bibfnamefont {C.~S.}\ \bibnamefont {Hwang}},\ }\href
  {https://doi.org/10.1038/srep15965} {\bibfield  {journal} {\bibinfo
  {journal} {Sci. Rep.}\ }\textbf {\bibinfo {volume} {5}},\ \bibinfo {pages}
  {15965} (\bibinfo {year} {2015})}\BibitemShut {NoStop}%
\bibitem [{\citenamefont {Heisig}\ \emph {et~al.}(2022)\citenamefont {Heisig},
  \citenamefont {Lange}, \citenamefont {Gutsche}, \citenamefont {Goß},
  \citenamefont {Hambsch}, \citenamefont {Locatelli}, \citenamefont {Menteş},
  \citenamefont {Genuzio}, \citenamefont {Menzel},\ and\ \citenamefont
  {Dittmann}}]{heisig_2022}%
  \BibitemOpen
  \bibfield  {author} {\bibinfo {author} {\bibfnamefont {T.}~\bibnamefont
  {Heisig}}, \bibinfo {author} {\bibfnamefont {K.}~\bibnamefont {Lange}},
  \bibinfo {author} {\bibfnamefont {A.}~\bibnamefont {Gutsche}}, \bibinfo
  {author} {\bibfnamefont {K.~T.}\ \bibnamefont {Goß}}, \bibinfo {author}
  {\bibfnamefont {S.}~\bibnamefont {Hambsch}}, \bibinfo {author} {\bibfnamefont
  {A.}~\bibnamefont {Locatelli}}, \bibinfo {author} {\bibfnamefont {T.~O.}\
  \bibnamefont {Menteş}}, \bibinfo {author} {\bibfnamefont {F.}~\bibnamefont
  {Genuzio}}, \bibinfo {author} {\bibfnamefont {S.}~\bibnamefont {Menzel}}, \
  and\ \bibinfo {author} {\bibfnamefont {R.}~\bibnamefont {Dittmann}},\ }\href
  {\doibase https://doi.org/10.1002/aelm.202100936} {\bibfield  {journal}
  {\bibinfo  {journal} {Adv. Electron. Mater.}\ }\textbf {\bibinfo {volume}
  {n/a}},\ \bibinfo {pages} {2100936} (\bibinfo {year} {2022})}\BibitemShut
  {NoStop}%
\bibitem [{\citenamefont {Hsu}\ \emph {et~al.}(2014)\citenamefont {Hsu},
  \citenamefont {Wang}, \citenamefont {Wan}, \citenamefont {Wang},
  \citenamefont {Chou}, \citenamefont {Lai}, \citenamefont {Lee},\ and\
  \citenamefont {Hou}}]{hsu_2014}%
  \BibitemOpen
  \bibfield  {author} {\bibinfo {author} {\bibfnamefont {C.-W.}\ \bibnamefont
  {Hsu}}, \bibinfo {author} {\bibfnamefont {Y.-F.}\ \bibnamefont {Wang}},
  \bibinfo {author} {\bibfnamefont {C.-C.}\ \bibnamefont {Wan}}, \bibinfo
  {author} {\bibfnamefont {I.-T.}\ \bibnamefont {Wang}}, \bibinfo {author}
  {\bibfnamefont {C.-T.}\ \bibnamefont {Chou}}, \bibinfo {author}
  {\bibfnamefont {W.-L.}\ \bibnamefont {Lai}}, \bibinfo {author} {\bibfnamefont
  {Y.-J.}\ \bibnamefont {Lee}}, \ and\ \bibinfo {author} {\bibfnamefont
  {T.-H.}\ \bibnamefont {Hou}},\ }\href {\doibase
  10.1088/0957-4484/25/16/165202} {\bibfield  {journal} {\bibinfo  {journal}
  {Nanotechnology}\ }\textbf {\bibinfo {volume} {25}},\ \bibinfo {pages}
  {165202} (\bibinfo {year} {2014})}\BibitemShut {NoStop}%
\bibitem [{\citenamefont {Yang}\ \emph {et~al.}(2012)\citenamefont {Yang},
  \citenamefont {Sheridan},\ and\ \citenamefont {Lu}}]{yang_2012}%
  \BibitemOpen
  \bibfield  {author} {\bibinfo {author} {\bibfnamefont {Y.}~\bibnamefont
  {Yang}}, \bibinfo {author} {\bibfnamefont {P.}~\bibnamefont {Sheridan}}, \
  and\ \bibinfo {author} {\bibfnamefont {W.}~\bibnamefont {Lu}},\ }\href
  {\doibase 10.1063/1.4719198} {\bibfield  {journal} {\bibinfo  {journal}
  {Appl. Phys. Lett.}\ }\textbf {\bibinfo {volume} {100}},\ \bibinfo {pages}
  {203112} (\bibinfo {year} {2012})}\BibitemShut {NoStop}%
\bibitem [{\citenamefont {Ferreyra}\ \emph
  {et~al.}(2020{\natexlab{a}})\citenamefont {Ferreyra}, \citenamefont
  {S{\'{a}}nchez}, \citenamefont {Aguirre}, \citenamefont {Acha}, \citenamefont
  {Bengi{\'{o}}}, \citenamefont {Lecourt}, \citenamefont {Lüders},\ and\
  \citenamefont {Rubi}}]{ferreyra_2020}%
  \BibitemOpen
  \bibfield  {author} {\bibinfo {author} {\bibfnamefont {C.}~\bibnamefont
  {Ferreyra}}, \bibinfo {author} {\bibfnamefont {M.~J.}\ \bibnamefont
  {S{\'{a}}nchez}}, \bibinfo {author} {\bibfnamefont {M.}~\bibnamefont
  {Aguirre}}, \bibinfo {author} {\bibfnamefont {C.}~\bibnamefont {Acha}},
  \bibinfo {author} {\bibfnamefont {S.}~\bibnamefont {Bengi{\'{o}}}}, \bibinfo
  {author} {\bibfnamefont {J.}~\bibnamefont {Lecourt}}, \bibinfo {author}
  {\bibfnamefont {U.}~\bibnamefont {Lüders}}, \ and\ \bibinfo {author}
  {\bibfnamefont {D.}~\bibnamefont {Rubi}},\ }\href {\doibase
  10.1088/1361-6528/ab6476} {\bibfield  {journal} {\bibinfo  {journal}
  {Nanotechnology}\ }\textbf {\bibinfo {volume} {31}},\ \bibinfo {pages}
  {155204} (\bibinfo {year} {2020}{\natexlab{a}})}\BibitemShut {NoStop}%
\bibitem [{\citenamefont {Nian}\ \emph {et~al.}(2007)\citenamefont {Nian},
  \citenamefont {Strozier}, \citenamefont {Wu}, \citenamefont {Chen},\ and\
  \citenamefont {Ignatiev}}]{nian_2007}%
  \BibitemOpen
  \bibfield  {author} {\bibinfo {author} {\bibfnamefont {Y.~B.}\ \bibnamefont
  {Nian}}, \bibinfo {author} {\bibfnamefont {J.}~\bibnamefont {Strozier}},
  \bibinfo {author} {\bibfnamefont {N.~J.}\ \bibnamefont {Wu}}, \bibinfo
  {author} {\bibfnamefont {X.}~\bibnamefont {Chen}}, \ and\ \bibinfo {author}
  {\bibfnamefont {A.}~\bibnamefont {Ignatiev}},\ }\href {\doibase
  10.1103/PhysRevLett.98.146403} {\bibfield  {journal} {\bibinfo  {journal}
  {Phys. Rev. Lett.}\ }\textbf {\bibinfo {volume} {98}},\ \bibinfo {pages}
  {146403} (\bibinfo {year} {2007})}\BibitemShut {NoStop}%
\bibitem [{\citenamefont {Rozenberg}\ \emph {et~al.}(2010)\citenamefont
  {Rozenberg}, \citenamefont {S\'anchez}, \citenamefont {Weht}, \citenamefont
  {Acha}, \citenamefont {Gomez-Marlasca},\ and\ \citenamefont
  {Levy}}]{rozenberg_2010}%
  \BibitemOpen
  \bibfield  {author} {\bibinfo {author} {\bibfnamefont {M.~J.}\ \bibnamefont
  {Rozenberg}}, \bibinfo {author} {\bibfnamefont {M.~J.}\ \bibnamefont
  {S\'anchez}}, \bibinfo {author} {\bibfnamefont {R.}~\bibnamefont {Weht}},
  \bibinfo {author} {\bibfnamefont {C.}~\bibnamefont {Acha}}, \bibinfo {author}
  {\bibfnamefont {F.}~\bibnamefont {Gomez-Marlasca}}, \ and\ \bibinfo {author}
  {\bibfnamefont {P.}~\bibnamefont {Levy}},\ }\href {\doibase
  10.1103/PhysRevB.81.115101} {\bibfield  {journal} {\bibinfo  {journal} {Phys.
  Rev. B}\ }\textbf {\bibinfo {volume} {81}},\ \bibinfo {pages} {115101}
  (\bibinfo {year} {2010})}\BibitemShut {NoStop}%
\bibitem [{\citenamefont {Zhang}\ \emph {et~al.}(2018)\citenamefont {Zhang},
  \citenamefont {Yoo}, \citenamefont {Menzel}, \citenamefont {Funck},
  \citenamefont {Cüppers}, \citenamefont {Wouters}, \citenamefont {Hwang},
  \citenamefont {Waser},\ and\ \citenamefont {Hoffmann-Eifert}}]{zhang_2018}%
  \BibitemOpen
  \bibfield  {author} {\bibinfo {author} {\bibfnamefont {H.}~\bibnamefont
  {Zhang}}, \bibinfo {author} {\bibfnamefont {S.}~\bibnamefont {Yoo}}, \bibinfo
  {author} {\bibfnamefont {S.}~\bibnamefont {Menzel}}, \bibinfo {author}
  {\bibfnamefont {C.}~\bibnamefont {Funck}}, \bibinfo {author} {\bibfnamefont
  {F.}~\bibnamefont {Cüppers}}, \bibinfo {author} {\bibfnamefont {D.~J.}\
  \bibnamefont {Wouters}}, \bibinfo {author} {\bibfnamefont {C.~S.}\
  \bibnamefont {Hwang}}, \bibinfo {author} {\bibfnamefont {R.}~\bibnamefont
  {Waser}}, \ and\ \bibinfo {author} {\bibfnamefont {S.}~\bibnamefont
  {Hoffmann-Eifert}},\ }\href {doi: 10.1021/acsami.8b09068} {\bibfield
  {journal} {\bibinfo  {journal} {ACS Appl. Mater. Interfaces}\ }\textbf
  {\bibinfo {volume} {10}},\ \bibinfo {pages} {29766} (\bibinfo {year}
  {2018})}\BibitemShut {NoStop}%
\bibitem [{\citenamefont {Siegel}\ \emph {et~al.}(2021)\citenamefont {Siegel},
  \citenamefont {Baeumer}, \citenamefont {Gutsche}, \citenamefont {von
  Witzleben}, \citenamefont {Waser}, \citenamefont {Menzel},\ and\
  \citenamefont {Dittmann}}]{siegel_2021}%
  \BibitemOpen
  \bibfield  {author} {\bibinfo {author} {\bibfnamefont {S.}~\bibnamefont
  {Siegel}}, \bibinfo {author} {\bibfnamefont {C.}~\bibnamefont {Baeumer}},
  \bibinfo {author} {\bibfnamefont {A.}~\bibnamefont {Gutsche}}, \bibinfo
  {author} {\bibfnamefont {M.}~\bibnamefont {von Witzleben}}, \bibinfo {author}
  {\bibfnamefont {R.}~\bibnamefont {Waser}}, \bibinfo {author} {\bibfnamefont
  {S.}~\bibnamefont {Menzel}}, \ and\ \bibinfo {author} {\bibfnamefont
  {R.}~\bibnamefont {Dittmann}},\ }\href {\doibase
  https://doi.org/10.1002/aelm.202000815} {\bibfield  {journal} {\bibinfo
  {journal} {Adv. Electron. Mater.}\ }\textbf {\bibinfo {volume} {7}},\
  \bibinfo {pages} {2000815} (\bibinfo {year} {2021})}\BibitemShut {NoStop}%
\bibitem [{\citenamefont {Cooper}\ \emph {et~al.}(2017)\citenamefont {Cooper},
  \citenamefont {Baeumer}, \citenamefont {Bernier}, \citenamefont {Marchewka},
  \citenamefont {La~Torre}, \citenamefont {Dunin-Borkowski}, \citenamefont
  {Menzel}, \citenamefont {Waser},\ and\ \citenamefont
  {Dittmann}}]{cooper_2017}%
  \BibitemOpen
  \bibfield  {author} {\bibinfo {author} {\bibfnamefont {D.}~\bibnamefont
  {Cooper}}, \bibinfo {author} {\bibfnamefont {C.}~\bibnamefont {Baeumer}},
  \bibinfo {author} {\bibfnamefont {N.}~\bibnamefont {Bernier}}, \bibinfo
  {author} {\bibfnamefont {A.}~\bibnamefont {Marchewka}}, \bibinfo {author}
  {\bibfnamefont {C.}~\bibnamefont {La~Torre}}, \bibinfo {author}
  {\bibfnamefont {R.~E.}\ \bibnamefont {Dunin-Borkowski}}, \bibinfo {author}
  {\bibfnamefont {S.}~\bibnamefont {Menzel}}, \bibinfo {author} {\bibfnamefont
  {R.}~\bibnamefont {Waser}}, \ and\ \bibinfo {author} {\bibfnamefont
  {R.}~\bibnamefont {Dittmann}},\ }\href {\doibase
  https://doi.org/10.1002/adma.201700212} {\bibfield  {journal} {\bibinfo
  {journal} {Adv. Mater.}\ }\textbf {\bibinfo {volume} {29}},\ \bibinfo {pages}
  {1700212} (\bibinfo {year} {2017})}\BibitemShut {NoStop}%
\bibitem [{\citenamefont {Cox}\ \emph {et~al.}(2021)\citenamefont {Cox},
  \citenamefont {Buckwell}, \citenamefont {Ng}, \citenamefont {Mannion},
  \citenamefont {Mehonic}, \citenamefont {Shearing}, \citenamefont {Fearn},\
  and\ \citenamefont {Kenyon}}]{cox_2021}%
  \BibitemOpen
  \bibfield  {author} {\bibinfo {author} {\bibfnamefont {H.~R.~J.}\
  \bibnamefont {Cox}}, \bibinfo {author} {\bibfnamefont {M.}~\bibnamefont
  {Buckwell}}, \bibinfo {author} {\bibfnamefont {W.~H.}\ \bibnamefont {Ng}},
  \bibinfo {author} {\bibfnamefont {D.~J.}\ \bibnamefont {Mannion}}, \bibinfo
  {author} {\bibfnamefont {A.}~\bibnamefont {Mehonic}}, \bibinfo {author}
  {\bibfnamefont {P.~R.}\ \bibnamefont {Shearing}}, \bibinfo {author}
  {\bibfnamefont {S.}~\bibnamefont {Fearn}}, \ and\ \bibinfo {author}
  {\bibfnamefont {A.~J.}\ \bibnamefont {Kenyon}},\ }\href {\doibase
  10.1063/5.0070046} {\bibfield  {journal} {\bibinfo  {journal} {APL Mater.}\
  }\textbf {\bibinfo {volume} {9}},\ \bibinfo {pages} {111109} (\bibinfo {year}
  {2021})}\BibitemShut {NoStop}%
\bibitem [{\citenamefont {Messerschmitt}\ \emph {et~al.}(2015)\citenamefont
  {Messerschmitt}, \citenamefont {Kubicek},\ and\ \citenamefont
  {Rupp}}]{messer_2015}%
  \BibitemOpen
  \bibfield  {author} {\bibinfo {author} {\bibfnamefont {F.}~\bibnamefont
  {Messerschmitt}}, \bibinfo {author} {\bibfnamefont {M.}~\bibnamefont
  {Kubicek}}, \ and\ \bibinfo {author} {\bibfnamefont {J.~L.~M.}\ \bibnamefont
  {Rupp}},\ }\href {\doibase https://doi.org/10.1002/adfm.201501517} {\bibfield
   {journal} {\bibinfo  {journal} {Adv. Funct. Mater.}\ }\textbf {\bibinfo
  {volume} {25}},\ \bibinfo {pages} {5117} (\bibinfo {year}
  {2015})}\BibitemShut {NoStop}%
\bibitem [{\citenamefont {Heisig}\ \emph {et~al.}(2018)\citenamefont {Heisig},
  \citenamefont {Baeumer}, \citenamefont {Gries}, \citenamefont {Mueller},
  \citenamefont {La~Torre}, \citenamefont {Luebben}, \citenamefont {Raab},
  \citenamefont {Du}, \citenamefont {Menzel}, \citenamefont {Mueller},
  \citenamefont {Jia}, \citenamefont {Mayer}, \citenamefont {Waser},
  \citenamefont {Valov}, \citenamefont {De~Souza},\ and\ \citenamefont
  {Dittmann}}]{heisig_2018}%
  \BibitemOpen
  \bibfield  {author} {\bibinfo {author} {\bibfnamefont {T.}~\bibnamefont
  {Heisig}}, \bibinfo {author} {\bibfnamefont {C.}~\bibnamefont {Baeumer}},
  \bibinfo {author} {\bibfnamefont {U.~N.}\ \bibnamefont {Gries}}, \bibinfo
  {author} {\bibfnamefont {M.~P.}\ \bibnamefont {Mueller}}, \bibinfo {author}
  {\bibfnamefont {C.}~\bibnamefont {La~Torre}}, \bibinfo {author}
  {\bibfnamefont {M.}~\bibnamefont {Luebben}}, \bibinfo {author} {\bibfnamefont
  {N.}~\bibnamefont {Raab}}, \bibinfo {author} {\bibfnamefont {H.}~\bibnamefont
  {Du}}, \bibinfo {author} {\bibfnamefont {S.}~\bibnamefont {Menzel}}, \bibinfo
  {author} {\bibfnamefont {D.~N.}\ \bibnamefont {Mueller}}, \bibinfo {author}
  {\bibfnamefont {C.-L.}\ \bibnamefont {Jia}}, \bibinfo {author} {\bibfnamefont
  {J.}~\bibnamefont {Mayer}}, \bibinfo {author} {\bibfnamefont
  {R.}~\bibnamefont {Waser}}, \bibinfo {author} {\bibfnamefont
  {I.}~\bibnamefont {Valov}}, \bibinfo {author} {\bibfnamefont {R.~A.}\
  \bibnamefont {De~Souza}}, \ and\ \bibinfo {author} {\bibfnamefont
  {R.}~\bibnamefont {Dittmann}},\ }\href {\doibase
  https://doi.org/10.1002/adma.201800957} {\bibfield  {journal} {\bibinfo
  {journal} {Adv. Mater.}\ }\textbf {\bibinfo {volume} {30}},\ \bibinfo {pages}
  {1800957} (\bibinfo {year} {2018})}\BibitemShut {NoStop}%
\bibitem [{\citenamefont {Kim}\ and\ \citenamefont {Choi}(2009)}]{5306148}%
  \BibitemOpen
  \bibfield  {author} {\bibinfo {author} {\bibfnamefont {S.}~\bibnamefont
  {Kim}}\ and\ \bibinfo {author} {\bibfnamefont {Y.-K.}\ \bibnamefont {Choi}},\
  }\href {\doibase 10.1109/TED.2009.2032597} {\bibfield  {journal} {\bibinfo
  {journal} {IEEE Trans. Electron Devices}\ }\textbf {\bibinfo {volume} {56}},\
  \bibinfo {pages} {3049} (\bibinfo {year} {2009})}\BibitemShut {NoStop}%
\bibitem [{\citenamefont {Tsuruoka}\ \emph {et~al.}(2016)\citenamefont
  {Tsuruoka}, \citenamefont {Valov}, \citenamefont {Mannequin}, \citenamefont
  {Hasegawa}, \citenamefont {Waser},\ and\ \citenamefont
  {Aono}}]{Tsuruoka_2016}%
  \BibitemOpen
  \bibfield  {author} {\bibinfo {author} {\bibfnamefont {T.}~\bibnamefont
  {Tsuruoka}}, \bibinfo {author} {\bibfnamefont {I.}~\bibnamefont {Valov}},
  \bibinfo {author} {\bibfnamefont {C.}~\bibnamefont {Mannequin}}, \bibinfo
  {author} {\bibfnamefont {T.}~\bibnamefont {Hasegawa}}, \bibinfo {author}
  {\bibfnamefont {R.}~\bibnamefont {Waser}}, \ and\ \bibinfo {author}
  {\bibfnamefont {M.}~\bibnamefont {Aono}},\ }\href {\doibase
  10.7567/jjap.55.06gj09} {\bibfield  {journal} {\bibinfo  {journal} {Jpn. J.
  Appl. Phys.}\ }\textbf {\bibinfo {volume} {55}},\ \bibinfo {pages} {06GJ09}
  (\bibinfo {year} {2016})}\BibitemShut {NoStop}%
\bibitem [{\citenamefont {Wang}\ \emph {et~al.}(2018)\citenamefont {Wang},
  \citenamefont {Ren}, \citenamefont {Zhang}, \citenamefont {Xiang},
  \citenamefont {Zhao}, \citenamefont {Zhou}, \citenamefont {Li}, \citenamefont
  {Wang}, \citenamefont {Zhang}, \citenamefont {Zhao}, \citenamefont {Fang},
  \citenamefont {Lu}, \citenamefont {Zhao}, \citenamefont {Zhao},\ and\
  \citenamefont {Yan}}]{wang_2018}%
  \BibitemOpen
  \bibfield  {author} {\bibinfo {author} {\bibfnamefont {J.}~\bibnamefont
  {Wang}}, \bibinfo {author} {\bibfnamefont {D.}~\bibnamefont {Ren}}, \bibinfo
  {author} {\bibfnamefont {Z.}~\bibnamefont {Zhang}}, \bibinfo {author}
  {\bibfnamefont {H.}~\bibnamefont {Xiang}}, \bibinfo {author} {\bibfnamefont
  {J.}~\bibnamefont {Zhao}}, \bibinfo {author} {\bibfnamefont {Z.}~\bibnamefont
  {Zhou}}, \bibinfo {author} {\bibfnamefont {X.}~\bibnamefont {Li}}, \bibinfo
  {author} {\bibfnamefont {H.}~\bibnamefont {Wang}}, \bibinfo {author}
  {\bibfnamefont {L.}~\bibnamefont {Zhang}}, \bibinfo {author} {\bibfnamefont
  {M.}~\bibnamefont {Zhao}}, \bibinfo {author} {\bibfnamefont {Y.}~\bibnamefont
  {Fang}}, \bibinfo {author} {\bibfnamefont {C.}~\bibnamefont {Lu}}, \bibinfo
  {author} {\bibfnamefont {C.}~\bibnamefont {Zhao}}, \bibinfo {author}
  {\bibfnamefont {C.}~\bibnamefont {Zhao}}, \ and\ \bibinfo {author}
  {\bibfnamefont {X.}~\bibnamefont {Yan}},\ }\href {\doibase 10.1063/1.5045649}
  {\bibfield  {journal} {\bibinfo  {journal} {Appl. Phys. Lett.}\ }\textbf
  {\bibinfo {volume} {113}},\ \bibinfo {pages} {122907} (\bibinfo {year}
  {2018})}\BibitemShut {NoStop}%
\bibitem [{\citenamefont {Gul}\ and\ \citenamefont {Efeoglu}(2022)}]{gul_2022}%
  \BibitemOpen
  \bibfield  {author} {\bibinfo {author} {\bibfnamefont {M.}~\bibnamefont
  {Gul}}\ and\ \bibinfo {author} {\bibfnamefont {H.}~\bibnamefont {Efeoglu}},\
  }\href {\doibase 10.1007/s10854-022-07864-z} {\bibfield  {journal} {\bibinfo
  {journal} {J. Mater. Sci.: Mater. Electron.}\ }\textbf {\bibinfo {volume}
  {33}},\ \bibinfo {pages} {7423 } (\bibinfo {year} {2022})}\BibitemShut
  {NoStop}%
\bibitem [{\citenamefont {Zurhelle}\ \emph {et~al.}(2022)\citenamefont
  {Zurhelle}, \citenamefont {Stehling}, \citenamefont {Waser}, \citenamefont
  {De~Souza},\ and\ \citenamefont {Menzel}}]{zurhelle_2022}%
  \BibitemOpen
  \bibfield  {author} {\bibinfo {author} {\bibfnamefont {A.~F.}\ \bibnamefont
  {Zurhelle}}, \bibinfo {author} {\bibfnamefont {W.}~\bibnamefont {Stehling}},
  \bibinfo {author} {\bibfnamefont {R.}~\bibnamefont {Waser}}, \bibinfo
  {author} {\bibfnamefont {R.~A.}\ \bibnamefont {De~Souza}}, \ and\ \bibinfo
  {author} {\bibfnamefont {S.}~\bibnamefont {Menzel}},\ }\href {\doibase
  https://doi.org/10.1002/admi.202101257} {\bibfield  {journal} {\bibinfo
  {journal} {Adv. Mater. Interfaces}\ }\textbf {\bibinfo {volume} {9}},\
  \bibinfo {pages} {2101257} (\bibinfo {year} {2022})}\BibitemShut {NoStop}%
\bibitem [{\citenamefont {Tsuchiya}\ \emph {et~al.}(2011)\citenamefont
  {Tsuchiya}, \citenamefont {Imai}, \citenamefont {Miyoshi}, \citenamefont
  {Glans}, \citenamefont {Guo},\ and\ \citenamefont
  {Yamaguchi}}]{tsuchiya_2011}%
  \BibitemOpen
  \bibfield  {author} {\bibinfo {author} {\bibfnamefont {T.}~\bibnamefont
  {Tsuchiya}}, \bibinfo {author} {\bibfnamefont {H.}~\bibnamefont {Imai}},
  \bibinfo {author} {\bibfnamefont {S.}~\bibnamefont {Miyoshi}}, \bibinfo
  {author} {\bibfnamefont {P.-A.}\ \bibnamefont {Glans}}, \bibinfo {author}
  {\bibfnamefont {J.}~\bibnamefont {Guo}}, \ and\ \bibinfo {author}
  {\bibfnamefont {S.}~\bibnamefont {Yamaguchi}},\ }\href {\doibase
  10.1039/C1CP21310E} {\bibfield  {journal} {\bibinfo  {journal} {Phys. Chem.
  Chem. Phys.}\ }\textbf {\bibinfo {volume} {13}},\ \bibinfo {pages} {17013}
  (\bibinfo {year} {2011})}\BibitemShut {NoStop}%
\bibitem [{\citenamefont {Mannequin}\ \emph {et~al.}(2016)\citenamefont
  {Mannequin}, \citenamefont {Tsuruoka}, \citenamefont {Hasegawa},\ and\
  \citenamefont {Aono}}]{manne_2016}%
  \BibitemOpen
  \bibfield  {author} {\bibinfo {author} {\bibfnamefont {C.}~\bibnamefont
  {Mannequin}}, \bibinfo {author} {\bibfnamefont {T.}~\bibnamefont {Tsuruoka}},
  \bibinfo {author} {\bibfnamefont {T.}~\bibnamefont {Hasegawa}}, \ and\
  \bibinfo {author} {\bibfnamefont {M.}~\bibnamefont {Aono}},\ }\href {\doibase
  https://doi.org/10.1016/j.apsusc.2016.04.099} {\bibfield  {journal} {\bibinfo
   {journal} {Appl. Surf. Sci.}\ }\textbf {\bibinfo {volume} {385}},\ \bibinfo
  {pages} {426} (\bibinfo {year} {2016})}\BibitemShut {NoStop}%
\bibitem [{\citenamefont {Carta}\ \emph {et~al.}(2016)\citenamefont {Carta},
  \citenamefont {Hitchcock}, \citenamefont {Guttmann}, \citenamefont {Regoutz},
  \citenamefont {Khiat}, \citenamefont {Serb}, \citenamefont {Gupta},\ and\
  \citenamefont {Prodromakis}}]{carta_2016}%
  \BibitemOpen
  \bibfield  {author} {\bibinfo {author} {\bibfnamefont {D.}~\bibnamefont
  {Carta}}, \bibinfo {author} {\bibfnamefont {A.~P.}\ \bibnamefont
  {Hitchcock}}, \bibinfo {author} {\bibfnamefont {P.}~\bibnamefont {Guttmann}},
  \bibinfo {author} {\bibfnamefont {A.}~\bibnamefont {Regoutz}}, \bibinfo
  {author} {\bibfnamefont {A.}~\bibnamefont {Khiat}}, \bibinfo {author}
  {\bibfnamefont {A.}~\bibnamefont {Serb}}, \bibinfo {author} {\bibfnamefont
  {I.}~\bibnamefont {Gupta}}, \ and\ \bibinfo {author} {\bibfnamefont
  {T.}~\bibnamefont {Prodromakis}},\ }\href {\doibase 10.1038/srep21525}
  {\bibfield  {journal} {\bibinfo  {journal} {Sci. Rep.}\ }\textbf {\bibinfo
  {volume} {6}},\ \bibinfo {pages} {21525} (\bibinfo {year}
  {2016})}\BibitemShut {NoStop}%
\bibitem [{\citenamefont {Lederer}\ \emph {et~al.}(2021)\citenamefont
  {Lederer}, \citenamefont {Abdulazhanov}, \citenamefont {Olivo}, \citenamefont
  {Lehninger}, \citenamefont {Kämpfe}, \citenamefont {Seidel},\ and\
  \citenamefont {Eng}}]{lederer_2021}%
  \BibitemOpen
  \bibfield  {author} {\bibinfo {author} {\bibfnamefont {M.}~\bibnamefont
  {Lederer}}, \bibinfo {author} {\bibfnamefont {S.}~\bibnamefont
  {Abdulazhanov}}, \bibinfo {author} {\bibfnamefont {R.}~\bibnamefont {Olivo}},
  \bibinfo {author} {\bibfnamefont {D.}~\bibnamefont {Lehninger}}, \bibinfo
  {author} {\bibfnamefont {T.}~\bibnamefont {Kämpfe}}, \bibinfo {author}
  {\bibfnamefont {K.}~\bibnamefont {Seidel}}, \ and\ \bibinfo {author}
  {\bibfnamefont {L.~M.}\ \bibnamefont {Eng}},\ }\href {\doibase
  10.1038/s41598-021-01724-2} {\bibfield  {journal} {\bibinfo  {journal} {Sci.
  Rep.}\ }\textbf {\bibinfo {volume} {11}},\ \bibinfo {pages} {22266} (\bibinfo
  {year} {2021})}\BibitemShut {NoStop}%
\bibitem [{\citenamefont {Ghenzi}\ \emph {et~al.}(2013)\citenamefont {Ghenzi},
  \citenamefont {S{\'{a}}nchez},\ and\ \citenamefont {Levy}}]{ghenzi_2013}%
  \BibitemOpen
  \bibfield  {author} {\bibinfo {author} {\bibfnamefont {N.}~\bibnamefont
  {Ghenzi}}, \bibinfo {author} {\bibfnamefont {M.~J.}\ \bibnamefont
  {S{\'{a}}nchez}}, \ and\ \bibinfo {author} {\bibfnamefont {P.}~\bibnamefont
  {Levy}},\ }\href {\doibase 10.1088/0022-3727/46/41/415101} {\bibfield
  {journal} {\bibinfo  {journal} {J. Phys. D: Appl. Phys.}\ }\textbf {\bibinfo
  {volume} {46}},\ \bibinfo {pages} {415101} (\bibinfo {year}
  {2013})}\BibitemShut {NoStop}%
\bibitem [{\citenamefont {Ghenzi}\ \emph {et~al.}(2014)\citenamefont {Ghenzi},
  \citenamefont {S\'anchez}, \citenamefont {Rubi}, \citenamefont {Rozenberg},
  \citenamefont {Urdaniz}, \citenamefont {Weissman},\ and\ \citenamefont
  {Levy}}]{ghenzi_2014}%
  \BibitemOpen
  \bibfield  {author} {\bibinfo {author} {\bibfnamefont {N.}~\bibnamefont
  {Ghenzi}}, \bibinfo {author} {\bibfnamefont {M.~J.}\ \bibnamefont
  {S\'anchez}}, \bibinfo {author} {\bibfnamefont {D.}~\bibnamefont {Rubi}},
  \bibinfo {author} {\bibfnamefont {M.~J.}\ \bibnamefont {Rozenberg}}, \bibinfo
  {author} {\bibfnamefont {C.}~\bibnamefont {Urdaniz}}, \bibinfo {author}
  {\bibfnamefont {M.}~\bibnamefont {Weissman}}, \ and\ \bibinfo {author}
  {\bibfnamefont {P.}~\bibnamefont {Levy}},\ }\href {\doibase
  10.1063/1.4875559} {\bibfield  {journal} {\bibinfo  {journal} {Appl. Phys.
  Lett.}\ }\textbf {\bibinfo {volume} {104}},\ \bibinfo {pages} {183505}
  (\bibinfo {year} {2014})}\BibitemShut {NoStop}%
\bibitem [{\citenamefont {Acevedo~Rom\'an}\ \emph {et~al.}(2017)\citenamefont
  {Acevedo~Rom\'an}, \citenamefont {Acha}, \citenamefont {Sanchez},
  \citenamefont {Levy},\ and\ \citenamefont {Rubi}}]{Acevedo_2017}%
  \BibitemOpen
  \bibfield  {author} {\bibinfo {author} {\bibfnamefont {W.}~\bibnamefont
  {Acevedo~Rom\'an}}, \bibinfo {author} {\bibfnamefont {C.}~\bibnamefont
  {Acha}}, \bibinfo {author} {\bibfnamefont {M.~J.}\ \bibnamefont {Sanchez}},
  \bibinfo {author} {\bibfnamefont {P.}~\bibnamefont {Levy}}, \ and\ \bibinfo
  {author} {\bibfnamefont {D.}~\bibnamefont {Rubi}},\ }\href@noop {} {\bibfield
   {journal} {\bibinfo  {journal} {Appl. Phys. Lett.}\ }\textbf {\bibinfo
  {volume} {110}},\ \bibinfo {pages} {053501} (\bibinfo {year}
  {2017})}\BibitemShut {NoStop}%
\bibitem [{\citenamefont {Ferreyra}\ \emph
  {et~al.}(2020{\natexlab{b}})\citenamefont {Ferreyra}, \citenamefont
  {Rengifo}, \citenamefont {S\'anchez}, \citenamefont {Everhardt},
  \citenamefont {Noheda},\ and\ \citenamefont {Rubi}}]{ferreyra_2020_2}%
  \BibitemOpen
  \bibfield  {author} {\bibinfo {author} {\bibfnamefont {C.}~\bibnamefont
  {Ferreyra}}, \bibinfo {author} {\bibfnamefont {M.}~\bibnamefont {Rengifo}},
  \bibinfo {author} {\bibfnamefont {M.}~\bibnamefont {S\'anchez}}, \bibinfo
  {author} {\bibfnamefont {A.}~\bibnamefont {Everhardt}}, \bibinfo {author}
  {\bibfnamefont {B.}~\bibnamefont {Noheda}}, \ and\ \bibinfo {author}
  {\bibfnamefont {D.}~\bibnamefont {Rubi}},\ }\href {\doibase
  10.1103/PhysRevApplied.14.044045} {\bibfield  {journal} {\bibinfo  {journal}
  {Phys. Rev. Appl.}\ }\textbf {\bibinfo {volume} {14}},\ \bibinfo {pages}
  {044045} (\bibinfo {year} {2020}{\natexlab{b}})}\BibitemShut {NoStop}%
\bibitem [{\citenamefont {Román~Acevedo}\ \emph {et~al.}(2020)\citenamefont
  {Román~Acevedo}, \citenamefont {van~den Bosch}, \citenamefont {Aguirre},
  \citenamefont {Acha}, \citenamefont {Cavallaro}, \citenamefont {Ferreyra},
  \citenamefont {Sánchez}, \citenamefont {Patrone}, \citenamefont {Aguadero},\
  and\ \citenamefont {Rubi}}]{acevedo_2020}%
  \BibitemOpen
  \bibfield  {author} {\bibinfo {author} {\bibfnamefont {W.}~\bibnamefont
  {Román~Acevedo}}, \bibinfo {author} {\bibfnamefont {C.~A.~M.}\ \bibnamefont
  {van~den Bosch}}, \bibinfo {author} {\bibfnamefont {M.~H.}\ \bibnamefont
  {Aguirre}}, \bibinfo {author} {\bibfnamefont {C.}~\bibnamefont {Acha}},
  \bibinfo {author} {\bibfnamefont {A.}~\bibnamefont {Cavallaro}}, \bibinfo
  {author} {\bibfnamefont {C.}~\bibnamefont {Ferreyra}}, \bibinfo {author}
  {\bibfnamefont {M.~J.}\ \bibnamefont {Sánchez}}, \bibinfo {author}
  {\bibfnamefont {L.}~\bibnamefont {Patrone}}, \bibinfo {author} {\bibfnamefont
  {A.}~\bibnamefont {Aguadero}}, \ and\ \bibinfo {author} {\bibfnamefont
  {D.}~\bibnamefont {Rubi}},\ }\href {\doibase 10.1063/1.5131854} {\bibfield
  {journal} {\bibinfo  {journal} {Applied Physics Letters}\ }\textbf {\bibinfo
  {volume} {116}},\ \bibinfo {pages} {063502} (\bibinfo {year}
  {2020})}\BibitemShut {NoStop}%
\bibitem [{\citenamefont {Román~Acevedo}\ \emph {et~al.}(2022)\citenamefont
  {Román~Acevedo}, \citenamefont {Aguirre}, \citenamefont {Ferreyra},
  \citenamefont {Sánchez}, \citenamefont {Rengifo}, \citenamefont {van~den
  Bosch}, \citenamefont {Aguadero}, \citenamefont {Noheda},\ and\ \citenamefont
  {Rubi}}]{acevedo_2022}%
  \BibitemOpen
  \bibfield  {author} {\bibinfo {author} {\bibfnamefont {W.}~\bibnamefont
  {Román~Acevedo}}, \bibinfo {author} {\bibfnamefont {M.~H.}\ \bibnamefont
  {Aguirre}}, \bibinfo {author} {\bibfnamefont {C.}~\bibnamefont {Ferreyra}},
  \bibinfo {author} {\bibfnamefont {M.~J.}\ \bibnamefont {Sánchez}}, \bibinfo
  {author} {\bibfnamefont {M.}~\bibnamefont {Rengifo}}, \bibinfo {author}
  {\bibfnamefont {C.~A.~M.}\ \bibnamefont {van~den Bosch}}, \bibinfo {author}
  {\bibfnamefont {A.}~\bibnamefont {Aguadero}}, \bibinfo {author}
  {\bibfnamefont {B.}~\bibnamefont {Noheda}}, \ and\ \bibinfo {author}
  {\bibfnamefont {D.}~\bibnamefont {Rubi}},\ }\href {\doibase
  10.1063/5.0073490} {\bibfield  {journal} {\bibinfo  {journal} {APL
  Materials}\ }\textbf {\bibinfo {volume} {10}},\ \bibinfo {pages} {011111}
  (\bibinfo {year} {2022})}\BibitemShut {NoStop}%
\bibitem [{\citenamefont {Ezhilvalavan}\ and\ \citenamefont
  {Tseng}(1999)}]{tseng_1999}%
  \BibitemOpen
  \bibfield  {author} {\bibinfo {author} {\bibfnamefont {S.}~\bibnamefont
  {Ezhilvalavan}}\ and\ \bibinfo {author} {\bibfnamefont {T.~Y.}\ \bibnamefont
  {Tseng}},\ }\href {\doibase 10.1023/A:1008970922635} {\bibfield  {journal}
  {\bibinfo  {journal} {J. Mater. Sci.: Mater. Electron.}\ }\textbf {\bibinfo
  {volume} {10}},\ \bibinfo {pages} {9 } (\bibinfo {year} {1999})}\BibitemShut
  {NoStop}%
\bibitem [{\citenamefont {Arif}\ \emph {et~al.}(2017)\citenamefont {Arif},
  \citenamefont {Balgis}, \citenamefont {Ogi}, \citenamefont {Iskandar},
  \citenamefont {Kinoshita}, \citenamefont {Nakamura},\ and\ \citenamefont
  {Okuyama}}]{arif_2017}%
  \BibitemOpen
  \bibfield  {author} {\bibinfo {author} {\bibfnamefont {A.~F.}\ \bibnamefont
  {Arif}}, \bibinfo {author} {\bibfnamefont {R.}~\bibnamefont {Balgis}},
  \bibinfo {author} {\bibfnamefont {T.}~\bibnamefont {Ogi}}, \bibinfo {author}
  {\bibfnamefont {F.}~\bibnamefont {Iskandar}}, \bibinfo {author}
  {\bibfnamefont {A.}~\bibnamefont {Kinoshita}}, \bibinfo {author}
  {\bibfnamefont {K.}~\bibnamefont {Nakamura}}, \ and\ \bibinfo {author}
  {\bibfnamefont {K.}~\bibnamefont {Okuyama}},\ }\href {\doibase
  10.1038/s41598-017-03509-y} {\bibfield  {journal} {\bibinfo  {journal} {Sci.
  Rep.}\ }\textbf {\bibinfo {volume} {7}},\ \bibinfo {pages} {3646} (\bibinfo
  {year} {2017})}\BibitemShut {NoStop}%
\bibitem [{\citenamefont {Kim}\ \emph {et~al.}(2016)\citenamefont {Kim},
  \citenamefont {Menzel}, \citenamefont {Wouters}, \citenamefont {Guo},
  \citenamefont {Robertson}, \citenamefont {Roesgen}, \citenamefont {Waser},\
  and\ \citenamefont {Rana}}]{kim_2016}%
  \BibitemOpen
  \bibfield  {author} {\bibinfo {author} {\bibfnamefont {W.}~\bibnamefont
  {Kim}}, \bibinfo {author} {\bibfnamefont {S.}~\bibnamefont {Menzel}},
  \bibinfo {author} {\bibfnamefont {D.~J.}\ \bibnamefont {Wouters}}, \bibinfo
  {author} {\bibfnamefont {Y.}~\bibnamefont {Guo}}, \bibinfo {author}
  {\bibfnamefont {J.}~\bibnamefont {Robertson}}, \bibinfo {author}
  {\bibfnamefont {B.}~\bibnamefont {Roesgen}}, \bibinfo {author} {\bibfnamefont
  {R.}~\bibnamefont {Waser}}, \ and\ \bibinfo {author} {\bibfnamefont
  {V.}~\bibnamefont {Rana}},\ }\href {\doibase 10.1039/C6NR03810G} {\bibfield
  {journal} {\bibinfo  {journal} {Nanoscale}\ }\textbf {\bibinfo {volume}
  {8}},\ \bibinfo {pages} {17774} (\bibinfo {year} {2016})}\BibitemShut
  {NoStop}%
\bibitem [{\citenamefont {Ma}\ \emph {et~al.}(2017)\citenamefont {Ma},
  \citenamefont {Wu}, \citenamefont {Wang},\ and\ \citenamefont
  {Dai}}]{ma_2017}%
  \BibitemOpen
  \bibfield  {author} {\bibinfo {author} {\bibfnamefont {X.}~\bibnamefont
  {Ma}}, \bibinfo {author} {\bibfnamefont {X.}~\bibnamefont {Wu}}, \bibinfo
  {author} {\bibfnamefont {Y.}~\bibnamefont {Wang}}, \ and\ \bibinfo {author}
  {\bibfnamefont {Y.}~\bibnamefont {Dai}},\ }\href {\doibase
  10.1039/C7CP03453A} {\bibfield  {journal} {\bibinfo  {journal} {Phys. Chem.
  Chem. Phys.}\ }\textbf {\bibinfo {volume} {19}},\ \bibinfo {pages} {18750}
  (\bibinfo {year} {2017})}\BibitemShut {NoStop}%
\bibitem [{\citenamefont {Gomez-Marlasca}\ \emph {et~al.}(2013)\citenamefont
  {Gomez-Marlasca}, \citenamefont {Ghenzi}, \citenamefont {Leyva},
  \citenamefont {Albornoz}, \citenamefont {Rubi}, \citenamefont {Stoliar},\
  and\ \citenamefont {Levy}}]{marlasca_2013}%
  \BibitemOpen
  \bibfield  {author} {\bibinfo {author} {\bibfnamefont {F.}~\bibnamefont
  {Gomez-Marlasca}}, \bibinfo {author} {\bibfnamefont {N.}~\bibnamefont
  {Ghenzi}}, \bibinfo {author} {\bibfnamefont {A.~G.}\ \bibnamefont {Leyva}},
  \bibinfo {author} {\bibfnamefont {C.}~\bibnamefont {Albornoz}}, \bibinfo
  {author} {\bibfnamefont {D.}~\bibnamefont {Rubi}}, \bibinfo {author}
  {\bibfnamefont {P.}~\bibnamefont {Stoliar}}, \ and\ \bibinfo {author}
  {\bibfnamefont {P.}~\bibnamefont {Levy}},\ }\href {\doibase
  10.1063/1.4800887} {\bibfield  {journal} {\bibinfo  {journal} {Journal of
  Applied Physics}\ }\textbf {\bibinfo {volume} {113}},\ \bibinfo {pages}
  {144510} (\bibinfo {year} {2013})}\BibitemShut {NoStop}%
\bibitem [{\citenamefont {Acha}(2017)}]{acha_2017}%
  \BibitemOpen
  \bibfield  {author} {\bibinfo {author} {\bibfnamefont {C.}~\bibnamefont
  {Acha}},\ }\href {\doibase 10.1063/1.4979723} {\bibfield  {journal} {\bibinfo
   {journal} {Journal of Applied Physics}\ }\textbf {\bibinfo {volume} {121}},\
  \bibinfo {pages} {134502} (\bibinfo {year} {2017})}\BibitemShut {NoStop}%
\bibitem [{\citenamefont {Murgatroyd}(1970)}]{Murga_1970}%
  \BibitemOpen
  \bibfield  {author} {\bibinfo {author} {\bibfnamefont {P.~N.}\ \bibnamefont
  {Murgatroyd}},\ }\href {\doibase 10.1088/0022-3727/3/2/308} {\bibfield
  {journal} {\bibinfo  {journal} {Journal of Physics D: Applied Physics}\
  }\textbf {\bibinfo {volume} {3}},\ \bibinfo {pages} {151} (\bibinfo {year}
  {1970})}\BibitemShut {NoStop}%
\bibitem [{Note1()}]{Note1}%
  \BibitemOpen
  \bibinfo {note} {We also tested the possible contributions of Schottky diodes
  present at both metal-oxide interfaces, finding that they don't significantly
  contribute to the electronic transport in the range of (small) voltages used
  to perform the fittings. They would only contribute with a small part of the
  conduction in the range of higher voltages, where some deviations between the
  fits and the experimental values can be observed.}\BibitemShut {Stop}%
\bibitem [{\citenamefont {Prezioso}\ \emph {et~al.}(2015)\citenamefont
  {Prezioso}, \citenamefont {Merrikh~Bayat}, \citenamefont {Hoskins},
  \citenamefont {Adam}, \citenamefont {Likharev},\ and\ \citenamefont
  {Strukov}}]{prez15}%
  \BibitemOpen
  \bibfield  {author} {\bibinfo {author} {\bibfnamefont {M.}~\bibnamefont
  {Prezioso}}, \bibinfo {author} {\bibfnamefont {F.}~\bibnamefont
  {Merrikh~Bayat}}, \bibinfo {author} {\bibfnamefont {B.~D.}\ \bibnamefont
  {Hoskins}}, \bibinfo {author} {\bibfnamefont {G.~C.}\ \bibnamefont {Adam}},
  \bibinfo {author} {\bibfnamefont {K.~K.}\ \bibnamefont {Likharev}}, \ and\
  \bibinfo {author} {\bibfnamefont {D.~B.}\ \bibnamefont {Strukov}},\
  }\href@noop {} {\bibfield  {journal} {\bibinfo  {journal} {Nature}\ }\textbf
  {\bibinfo {volume} {521}},\ \bibinfo {pages} {31} (\bibinfo {year}
  {2015})}\BibitemShut {NoStop}%
\end{thebibliography}%

\end{document}